\documentclass[12pt]{article}
\usepackage{epsfig}
\usepackage{lscape}
\usepackage{longtable}

\begin{document}
\begin{center}

{\Large {\bf Radioactive contamination of scintillators}}

\vskip 1.6cm

{\bf F.A.~Danevich$^{a,b}$, V.I.~Tretyak$^{a}$}

\vskip 0.4cm

$^{a}${\it Institute for Nuclear Research, Kyiv 03028, Ukraine}

$^{b}${\it CSNSM, Univ. Paris-Sud, CNRS/IN2P3, Univ. Paris-Saclay,
91405 Orsay, France}

\end{center}

\vskip 1.0cm

\begin{abstract}
Low counting experiments (search for double $\beta$ decay and dark
matter particles, measurements of neutrino fluxes from different
sources, search for hypothetical nuclear and subnuclear processes,
low background $\alpha$, $\beta$, $\gamma$ spectrometry) require
extremely low background of a detector. Scintillators are widely
used to search for rare events both as conventional scintillation
detectors and as cryogenic scintillating bolometers. Radioactive
contamination of a scintillation material plays a key role to
reach low level of background. Origin and nature of radioactive
contamination of scintillators, experimental methods and results
are reviewed. A programme to develop radiopure crystal
scintillators for low counting experiments is discussed briefly.
\end{abstract}

\vskip 0.3cm


\vskip 0.3cm


\vskip 0.8cm

\section{Introduction}
\label{sec:intro}

Search for neutrinoless double beta decay, dark matter particles,
measurements of solar and reactor neutrino fluxes, tests of
fundamental laws with increasing accuracy (e.g. Pauli exclusion
principle, charge conservation, stability of nucleons, etc.),
search for hypothetical particles and effects beyond the Standard
Model (axions; charge, magnetic momentum, mass of neutrino; etc.) are
the topics of astroparticle physics. Scintillation detectors
possess a range of important properties for low counting
experiments: presence of certain chemical elements (important in
searches for double decay, investigations of rare $\beta$ and
$\alpha$ decays) or variety of elements (which can be exploited in
dark matter detectors to probe new areas of parameter space),
large sensitive volume, reasonable energy resolution (very high in
a case of cryogenic scintillating bolometers), low energy
threshold, long time stability of operation, pulse-shape
discrimination ability, low cost.

Radiopurity is a crucial property of scintillation material to
reach low background counting rate of a detector. We review here
application of scintillation materials in astroparticle physics,
experimental methods to measure activities of radionuclides inside
scintillators, origin and nature of radioactive contamination,
radiopurity data for different scintillation materials. A programme to develop
radiopure crystal scintillators for low counting experiments is
discussed briefly. We do not consider here noble gas based low
background scintillation detectors referring readers to the
reviews \cite{Bernabei:2008a,Bernabei:2015a,Aprile:2010}.

\section{Scintillators in astroparticle physics}
\label{sec:sc-app}

\subsection{Double $\beta$ decay}
\label{sec:sc-app-2b}

Observations of neutrino oscillations give clear evidence that
neutrino is a massive particle (see e.g.
\cite{Mohapatra:2007,Fogli:2012,Gonzalez:2016,Vissani:2017}). The neutrinoless
double beta ($0\nu2\beta$) decay violates the lepton number
conservation and is possible if neutrino is a Majorana particle
(identical to its antiparticle) with non-zero mass
\cite{Schechter:1982}. The process is one of the ways to
investigate properties of neutrino and weak interaction, and to
test the Standard Model of particle physics
\cite{Barrea:2012,Rodejohann:2012,Deppisch:2012,Bilenky:2015,Delloro:2016,Vergados:2016}.
In particular, investigation of this phenomenon could allow to
determine the absolute scale of the neutrino mass and the neutrino
mass hierarchy. The $0\nu2\beta$ decay is still not observed
despite the seventy years of searches (for the status of
$0\nu2\beta$ decay searches we refer reader to the reviews
\cite{Elliott:2012,Giuliani:2012,Cremonesi:2014,Gomes:2015,Sarazin:2015,Pas:2015}
and recent experiments
\cite{EXO-200,NEMO-3,CUORE,GERDA,Gando:2017}).

High sensitivity experiments to search for double $\beta$
processes in different nuclei are strongly required both for
theoretical and experimental reasons. Despite valuable theoretical
efforts, there is a substantial difference in the $0\nu2\beta$
nuclear matrix elements (NME) calculated by using different
nuclear models \cite{Engel:2017}. Therefore, the theory does not
provide suggestions of a nucleus with the highest decay
probability. From the experimental point of view, since
$0\nu2\beta$ decay is expected to be an extremely rare process,
detection of the decay in one nucleus naturally will call for
support of the observation with other nuclei. Furthermore,
development of experimental methods for different nuclei is
required taking into account possible breakthroughs in
experimental techniques \cite{Giuliani:2018a}. The recent progress
in low-temperature scintillating bolometers is a good example.

There are quite a big number of scintillation materials which
contain double $\beta$ active isotopes. For this reason
scintillators are widely used in double $\beta$ decay experiments.
It is worth to mention a pioneer work of der Mateosian and
Goldhaber to search for neutrinoless double $\beta$ decay of
$^{48}$Ca by using enriched and depleted in $^{48}$Ca calcium
fluoride [CaF$_2$(Eu)] crystal scintillators
\cite{Mateosian:1966}. During the last two decades several high
sensitivity studies of the double $\beta$ decay processes were
performed using crystal scintillators with specific candidate
nuclei. The most sensitive $2\beta$ experiments with crystal
scintillators are reported in Table \ref{tab1}.

\clearpage
 \begin{landscape}
\begin{table*}[htbp]
\caption{The most sensitive calorimetric $2\beta$ experiments with
scintillators. Half-life limits are given at 90\% confidence
level.}
\begin{center}
\resizebox{1.0\textwidth}{!}{
\begin{tabular}{llll}
 \hline
 $2\beta$ transition                & Scintillator          & Main results: half-life (channels)                                                        & Reference \\
  ~                                 & ~                     & ~                                                                                         & ~         \\
  \hline
 $^{40}$Ca$\rightarrow$$^{40}$Ar    & CaWO$_4$              & $\geq 9.9\times10^{21}$ yr ($2\nu2K$)                                                     & \cite{Angloher:2016a} \\
 ~                                  & CaWO$_4$              & $\geq 1.4\times10^{22}$ yr ($0\nu2\varepsilon$)                                           & \cite{Angloher:2016a} \\
   \hline
 $^{46}$Ca$\rightarrow$$^{46}$Ti    & CaF$_2$(Eu)           & $\geq 1.0\times10^{17}$ yr ($0\nu2\beta^-$)                                               & \cite{Belli:1999a} \\
  \hline
 $^{48}$Ca$\rightarrow$$^{48}$Ti    & CaF$_2$(Eu)           & $\geq 5.8\times10^{22}$ yr ($0\nu2\beta^-$)                                               & \cite{Umehara:2008} \\
  \hline
 $^{70}$Zn$\rightarrow$$^{70}$Ge    & ZnWO$_4$              & $\geq 3.8\times10^{18}$ yr ($2\nu2\beta^-$)                                               & \cite{Belli:2011a} \\
  ~                                 & ZnWO$_4$              & $\geq 3.2\times10^{19}$ yr ($0\nu2\beta^-$)                                               & \cite{Belli:2011a} \\
 \hline
 $^{64}$Zn$\rightarrow$$^{64}$Ni    & ZnWO$_4$              & $\geq 1.1\times10^{19}$ yr ($2\nu2K$)                                                     & \cite{Belli:2011a} \\
  ~                                 & ZnWO$_4$              & $\geq 9.4\times10^{20}$ yr ($2\nu\varepsilon\beta^+$)                                     & \cite{Belli:2011a} \\
\hline
$^{82}$Se$\rightarrow$$^{82}$Kr    & Zn$^{82}$Se            & $\geq 2.4\times10^{24}$ yr
($0\nu2\beta^-$)                                            & \cite{Azzolini:2018} \\
  \hline
 $^{100}$Mo$\rightarrow$$^{100}$Ru  & ZnMoO$_4$             & $=[7.15 \pm 0.37\textrm{(stat.)} \pm 0.66\textrm{(syst.)}]\times 10^{18}$ yr ($2\nu2\beta^-$) & \cite{Cardani:2014} \\
 ~                                  & Li$_2$$^{100}$MoO$_4$ & $=[6.90 \pm 0.15\textrm{(stat.)}\pm 0.37\textrm{(syst.)}]\times 10^{18}$ yr ($2\nu2\beta^-$)  & \cite{Armengaud:2017a} \\
 ~                                  & Li$_2$$^{100}$MoO$_4$ & $\geq 7.0\times10^{22}$ yr ($0\nu2\beta^-$)                                               & \cite{Poda:2017a} \\
 ~                                  & $^{48\textrm{depl}}$Ca$^{100}$MoO$_4$ & $\geq 4.0\times10^{21}$ yr ($0\nu2\beta^-$)                                          & \cite{So:2012a} \\
 \hline
 $^{106}$Cd$\rightarrow$$^{106}$Pd  & $^{106}$CdWO$_4$      & $\geq 1.1\times10^{21}$ yr ($2\nu\varepsilon\beta^+$)                                     & \cite{Belli:2016a} \\
 ~                                  & $^{106}$CdWO$_4$      & $\geq 2.2\times10^{21}$ yr ($0\nu\varepsilon\beta^+$)                                     & \cite{Belli:2012a} \\

 \hline
 $^{108}$Cd$\rightarrow$$^{108}$Pd  & CdWO$_4$              & $\geq 1.1\times10^{18}$ yr ($2\nu2K$)                                                     & \cite{Belli:2008a} \\
 ~                                  & CdWO$_4$              & $\geq 1.0\times10^{18}$ yr ($0\nu2\varepsilon$)                                           & \cite{Belli:2008a} \\
 \hline
 $^{114}$Cd$\rightarrow$$^{114}$Sn  & CdWO$_4$              & $\geq 1.3\times10^{18}$ yr ($2\nu2\beta^-$)                                               & \cite{Belli:2008a} \\
 ~                                  & CdWO$_4$              & $\geq 1.1\times10^{21}$ yr ($0\nu2\beta^-$)                                               & \cite{Belli:2008a} \\
 \hline
 $^{116}$Cd$\rightarrow$$^{116}$Sn  & $^{116}$CdWO$_4$      & $=[2.69 \pm 0.02\textrm{(stat.)}\pm 0.14\textrm{(syst.)}]\times 10^{19}$ yr ($2\nu2\beta^-$)    & \cite{Polischuk:2017a} \\
 ~                                  & $^{116}$CdWO$_4$      & $\geq 2.4\times10^{23}$ yr ($0\nu2\beta^-$)                                                 & \cite{Polischuk:2017a} \\
 \hline
 $^{136}$Xe$\rightarrow$$^{136}$Ba  & Xenon-loaded          & $=[2.21 \pm 0.02\textrm{(stat.)}\pm 0.07\textrm{(syst.)}]\times 10^{21}$ yr ($2\nu2\beta^-$)    & \cite{Gando:2017} \\
 ~                                  & liquid scintillator   & $\geq 1.07\times10^{26}$ yr ($0\nu2\beta^-$)                                                 & \cite{Gando:2017} \\
 \hline
 $^{130}$Ba$\rightarrow$$^{130}$Xe  & BaF$_2$               & $\geq 1.4\times10^{17}$ yr ($0\nu\varepsilon\beta^+$)                                     & \cite{Cerulli:2004} \\
 \hline
 $^{136}$Ce$\rightarrow$$^{136}$Ba  & CeCl$_3$              & $\geq 3.2\times10^{16}$ yr ($2\nu2K$)                                                    & \cite{Belli:2011b} \\
 \hline
 $^{160}$Gd$\rightarrow$$^{160}$Dy  & Gd$_2$SiO$_5$(Ce)     & $\geq 1.9\times10^{19}$ yr ($2\nu2\beta^-$)                                               & \cite{Danevich:2001a} \\
 ~                                  & Gd$_2$SiO$_5$(Ce)     & $\geq 1.3\times10^{21}$ yr ($0\nu2\beta^-$)                                               & \cite{Danevich:2001a} \\
 \hline
 $^{180}$W$\rightarrow$$^{180}$Hf   & CaWO$_4$              & $\geq 3.1\times10^{19}$ yr ($2\nu2K$)                                                     & \cite{Angloher:2016a} \\
 ~                                  & CaWO$_4$              & $\geq 9.4\times10^{18}$ yr ($0\nu2\varepsilon$)                                           & \cite{Angloher:2016a} \\
 \hline
 $^{186}$W$\rightarrow$$^{186}$Os   & ZnWO$_4$              & $\geq 2.3\times10^{19}$ yr ($2\nu2\beta^-$)                                               & \cite{Belli:2011a} \\
 ~                                  & $^{116}$CdWO$_4$      & $\geq 1.1\times10^{21}$ yr ($0\nu2\beta^-$)                                               & \cite{Danevich:2003a} \\
 \hline
 \end{tabular}
 }
 \end{center}
 \label{tab1}
 \end{table*}
\end{landscape}

Large-scale scintillator-based projects to search for neutrinoless
double $\beta$ decay with sensitivity on the level of the inverted
hierarchy of the neutrino masses have been proposed
\cite{Chen:2005a,Lozza:2016a,Yoshida:2005a}. In this regard, it is
of note that the SuperNEMO double $\beta$ decay project
\cite{SuperNEMO} intents to utilize a large amount of plastic
scintillators for the calorimeter of the detector
\cite{Barabash:2017a}.

Cryogenic scintillation bolometers (see e.g.
\cite{Alessandrello:1998a,Pirro:2006a,Gorla:2008a,Pirro:2017a,Poda:2017b}),
with a typical energy resolution of a few keV and potentially with
strong reduced background, look perspective technique for future
$0\nu2\beta$ decay experiments able to explore the full range of
the inverted hierarchy of the neutrino mass (half-life sensitivity
on the level of $10^{26}-10^{28}$ years)
\cite{AMoRE,CUPID,CUPID-RD}. At present ZnSe
\cite{Arnaboldi:2011a,Beeman:2013a,Casali:2017a,Azzolini:2018},
Li$_2$MoO$_4$ \cite{Bekker:2016a,Armengaud:2017a} and CaMoO$_4$
\cite{Lee:2011a,Kim:2015a}, CdWO$_4$
\cite{Arnaboldi:2010a,Artusa:2014a,Barabash:2016a} are considered
as the most promising materials for high sensitivity
scintillating-bolometers $2\beta$ decay experiments. The
large-scale experiments intend to use $10^{2}-10^{3}$ kg of highly
radiopure scintillators enriched in the isotopes of interest. It
should be mentioned R\&D of other crystal scintillators containing
molybdenum: ZnMoO$_4$
\cite{Beeman:2012a,Beeman:2012b,Armengaud:2017a},
Li$_2$Zn$_2$(MoO$_4$)$_3$ \cite{Bashmakova:2009},
Li$_2$Mg$_2$(MoO$_4$)$_3$ \cite{Danevich:2018a}, Na$_2$Mo$_2$O$_7$
\cite{Spassky:2017}, Sr$_2$MoO$_4$ \cite{Mikhailik:2015}.

The double beta decay projects require as much as possible low, in
ideal case zero, background of a detector in a region of
interest\footnote{The decay energy of the most promising $2\beta$
nuclei is $\sim 2-3$ MeV.}. The most dangerous radionuclides for
$2\beta$ experiments are $^{226}$Ra and $^{228}$Th, since their
daughters ($^{214}$Bi and $^{208}$Tl) have large energies of
decay: $Q_{\beta}=3270$ keV and $Q_{\beta}=4999$ keV,
respectively. Potassium typically contributes to the energies
below 1461 keV. However, $^{40}$K can produce background hampering
$2\nu2\beta$ measurements. Presence of cosmogenic radioactivity
should be also controlled and decreased as much as possible. A
reachable (and measurable with present instrumentation) level of a
few $\mu$Bq/kg in crystal scintillators is discussed now (see,
e.g.
\cite{Beeman:2012a,Armengaud:2017a,Zuber:2007a,Umehara:2015,Luqman:2017a}).

\subsection{Dark matter}
\label{sec:sc-app-DM}

There is an evidence for a large amount of invisible (dark) matter
in the Universe which reveals itself only through gravitational
interaction. Weakly interacting massive particles (WIMPs), in
particular neutralino, predicted by the Minimal Supersymmetric
extensions of the Standard Model, are one of the many possible
candidates of dark matter
\cite{Bertone:2005a,Steffen:2009,Feng:2010,Bramante:2016a,Young:2017}.
In direct detection investigation the annual modulation signature
is a powerful tool because it is independent on the nature of the
dark matter candidate. It is exploited e.g. by DAMA
\cite{Bernabei:2010a} with highly radiopure NaI(Tl) crystal
scintillators.

In the case of the WIMP scenarios, WIMPs can be detected due to
their scattering on nuclei producing low energy nuclear recoils.
An extremely low counting rate (less than several counts kg$^{-1}$
d$^{-1}$) and small energy of recoil nuclei (below ~100 keV) are
expected in experiments to search for the WIMPs. Direct methods of
WIMP detection are based on registration of ionization,
scintillation or heat release caused by recoil nucleus embedded in
the material of the detector; the nucleus could be in an excited
state. At present, most sensitive direct experiments apply a
variety of detection techniques for WIMP search: semiconductor
detectors
\cite{Baudis:2000a,Morales:2002a,Aalseth:2008a,Lin:2009a,CDEX-1},
noble gases based detectors
\cite{Lebedenko:2009a,Angle:2010a,Benetti:2008a,XMASS,XENON1T,PandaX-II,LUX,ArDM},
bubble chambers \cite{PICO}, cryogenic bolometers
\cite{Sanglard:2005a,Ahmed:2009a,Agnese:2016a}. Crystal
scintillators are applied in conventional scintillation detectors
\cite{Bernabei:2000a,Alner:2005a,Kim:2012a,Bernabei:2010a,Belli:2014}
and in cryogenic scintillating bolometers, that use simultaneous
registration of heat and light signals from crystal scintillators
to reject background caused by electrons
\cite{CRESST-II,CRESST-III}. There are several dark matter
experiments in preparation using sodium iodine crystal
scintillators as conventional room temperature scintillation
detectors \cite{SABRE,ANAIS,PICOLON,DM-Ice17,COSINE-100} and
low-temperature scintillating bolometers \cite{COSINUS}.
Utilization of undoped CsI crystal scintillator as low-temperature
scintillating bolometer is considered in
Ref.\cite{Angloher:2016b}. High scintillation efficiency CaI$_2$
crystal scintillators (absolute light output $\sim106,000$
photon/MeV) were recently proposed as WIMP detectors aiming at
decreasing the energy threshold \cite{CaI2-TAPU2017,Kamada:2017}.
Anisotropic ZnWO$_4$ crystal scintillator is proposed to search
for directionality of dark matter signals
\cite{Capella:2013a,Cerulli:2017a}.

Radioactive contamination of target scintillation crystals plays a
key role to decrease background in the experiments. Counting rate
of a few counts kg$^{-1}$ d$^{-1}$ in the energy interval up to
$\sim20$ keV is typical in the present scintillator-based dark
matter experiments \cite{Kim:2012a,Bernabei:2010a,CRESST-III}. The
radioactive contamination of crystal scintillators used in dark
matter experiments by potassium, uranium, radium, thorium and
their daughters limits the experiments sensitivity. Besides,
primordial ($^{87}$Rb, $^{113}$Cd, $^{115}$In, $^{138}$La,
$^{176}$Lu, $^{187}$Re), cosmogenic (we refer reader to the review
\cite{Cebrian:2017a}) and artificial ($^{60}$Co, $^{134}$Cs,
$^{137}$Cs, etc.) $\beta$ active nuclides can produce background in
scintillation dark matter experiments too. It should be stressed
that presence of these radionuclides in crystals on the levels
significant for dark matter experiments can be detected in
practice only under extremely low background conditions with high
detection efficiency (actually, in the course of dark matter
experiments).

\subsection{Measurements of neutrino fluxes}
\label{sec:sc-app-nu-flux}

Measurements of solar and reactor neutrino fluxes allow to refine
our understanding of neutrino properties, in particular, to
determine parameters of the Pontecorvo-Maki-Nakagawa-Sakata matrix
and to test the Mikheyev-Smirnov-Wolfenstein effect
\cite{Fogli:2012,Capozzi:2016a,Qiana:2015a,Salas:2017a}. These
experiments require large-volume detectors (tens - thousands tons)
with an extremely low level of radioactive contamination
($\sim$nBq/kg). Especially purified liquid scintillators are used
for the real-time measurements of the solar neutrino flux in the
Borexino experiment \cite{Smirnov:2015a} and to measure
anti-neutrino flux from distant nuclear reactors in the KamLAND
detector \cite{Abe:2008a}.

Low radioactive liquid scintillation detectors are used in
experiments to measure antineutrino fluxes from nuclear reactors
\cite{Vogel:2015a}. There are several projects of large scale
experiments with reactor neutrinos. The Daya Bay \cite{An:2012a},
Double Chooz \cite{Abe:2016a}, and RENO \cite{Choi:2016a}
experiments utilized gadolinium-doped liquid scintillator to
measure the value of the neutrino mixing angle $\theta_{13}$. The
ambitious JUNO project intents to utilize 20 kton liquid
scintillation underground detector aiming at determination of the
neutrino mass hierarchy \cite{JUNO:2016,JUNO:2017}. It is expected
that the intrinsic radiopurity of the scintillator should be
better \cite{JUNO:2015} to those reached in the Borexino and
KamLAND detectors.

\subsection{Search for hypothetical processes and particles}
\label{sec:sc-app-hypo}

Scintillation detectors were used in a number of experiments to
search for hypothetical processes beyond the Standard Model: decay
of electron with violation of electric charge conservation, decay
of nucleons and pairs of nucleons, charge non-conserving (CNC)
beta decay, violation of the Pauli exclusion principle (PEP),
search for magnetic momentum of neutrino, etc. Searches for the
PEP violation were realized in
\cite{Ejiri:1992a,Belli:1999b,Bernabei:2009a} utilizing data of
low background experiments with NaI(Tl) crystal scintillators.
Data of the DAMA experiment have also been used to search for CNC
transitions in $^{23}$Na and $^{127}$I nuclei \cite{Belli:1999c},
searches for nucleons decay to invisible channels
\cite{Bernabei:2000b}, instability of electron \cite{Belli:1999b,Belli:2001a},
solar axions \cite{Belli:2001b}. Large mass and ultra-low level of
radioactive background of the Borexino detector have allowed to
establish new limits on processes of decay of nucleons into
invisible channels (for example, with emission of only neutrinos)
\cite{Back:2003a}, on magnetic momentum of neutrino
\cite{Agostini:2017a}, to search for solar axions
\cite{Bellini:2008a,Bellini:2012a} and for the PEP violation
\cite{Bellini:2010a}.
Invisible nucleons decays were searched for also in the SNO and
KamLAND experiments \cite{Ahmed:2004,Araki:2006}.
In work \cite{Bernabei:2006a} search for
activity with electric charge non-conservation was realized in the
experiment with LaCl$_3$(Ce) crystal scintillator.

\subsection{Investigation of rare $\alpha$ and $\beta$ decays}
\label{sec:sc-app-rare-a-b}

Crystal scintillators, both as ordinary scintillation detectors
and as cryogenic scintillating bolometers, are successfully used
to investigate rare $\alpha$ and $\beta$ decays. The half-life and
the spectrum shape of the fourth-forbidden $\beta$ decay of
$^{113}$Cd were measured with the help of CdWO$_4$ crystal
scintillators
\cite{Alessandrello:1994a,Danevich:1996a,Belli:2007a}. In the
experiment \cite{Belli:2007a} the $^{113}$Cd half-life was
determined with the highest to-date accuracy $T_{1/2} = (8.04\pm
0.05)\times 10^{15}$ yr. Liquid scintillators were used to measure
shape of spectrum and half-life of long-lived isotopes $^{87}$Rb
($T_{1/2} = 4.967(32)\times10^{10}$ yr \cite{Kossert:2003a}) and
$^{115}$In ($T_{1/2} = 4.41(26)\times 10^{14}$ yr
\cite{Pfeiffer:1979a}). Recently YVO$_4$ scintillating bolometer
was identified as a promising tool for investigation of rare
($T_{1/2} \simeq10^{17}$ yr) $\beta^-$ and EC decays of $^{50}$V
\cite{Pattavina:2018}. In 1960 Beard and Kelly have utilized small
CaWO$_4$ and CdWO$_4$ crystals in low background experiments to
search for alpha activity of natural tungsten \cite{Beard:1960a}.
However, even more sensitive experiment with CdWO$_4$ crystal
scintillator did not allow to observe $\alpha$ decays of tungsten
\cite{Georgadze:1995a}. The first indication on the decay of
$^{180}$W with a half-life $T_{1/2} = 1.1\times10^{18}$ yr was
obtained with the help of enriched $^{116}$CdWO$_4$ crystal
scintillator \cite{Danevich:2003b}. This observation was confirmed
with CaWO$_4$ crystals (as cryogenic scintillating bolometer
\cite{Cozzini:2004a} and room temperature scintillation detector
\cite{Zdesenko:2005a}) and ZnWO$_4$ crystal scintillator
\cite{Belli:2011c}. The alpha activity of bismuth ($^{209}$Bi,
considered before as the heaviest stable element in the nature)
with the half-life $T_{1/2}=(1.9\pm0.2)\times 10^{19}$ yr has been
detected with the help of BGO cryogenic scintillator
\cite{Marcillac:2003a}. In the same approach, decay of $^{209}$Bi
to the first excited level of $^{205}$Tl was also observed
\cite{Beeman:2012c}. An indication on rare $\alpha$ activity of
natural europium ($^{151}$Eu) obtained in the low background
experiment by using CaF$_2$(Eu) crystal scintillator
\cite{Belli:2007b} was confirmed with Li$_6$Eu(BO$_3$)$_3$ crystal
scintillator operated as a scintillating bolometer
\cite{Casali:2014a}. Half-life limits on $\alpha$ decays of Pb
isotopes were obtained with PbWO$_4$ scintillating bolometer grown
from ancient Roman lead with low $^{210}$Pb activity
\cite{Beeman:2013b}. ZnWO$_4$ scintillating bolometer doped with
enriched $^{148}$Sm was used for more precise measurement of
$^{148}$Sm half-life as $T_{1/2} = 6.4^{+1.2}_{-1.3}\times
10^{15}$ yr \cite{Casali:2016a}. An interesting study of the L/K
electron capture ratio in the decay of $^{207}$Bi decay to the
1633 keV excited level of $^{207}$Pb was realized with a BGO
scintillating bolometer contaminated by $^{207}$Bi
\cite{Coron:2012a}. Rare (probability of $\sim10^{-9}$) emission
of e$^+$e$^-$ pairs in $\alpha$ decay of $^{241}$Am was measured
with NaI(Tl) scintillators in Ref. \cite{Bernabei:2013}.

\section{Experimental methods to measure radioactive contamination of
scintillators} \label{sec:exp-meth}

Methods of determination of radioactive contaminants in
scintillators could be classified as direct, when characteristic
radioactivity of specific isotope is detected, and indirect, which
give quantitative conclusion on the presence of specific isotope
on the basis of measurements of contamination by the corresponding
chemical element (by mass-spectrometry or fluorescence methods) or
its daughters (by neutron activation analysis). The main
characteristics of elements containing primordial radioactive
isotopes are presented in Table \ref{tab2}. While indirect methods
can be applied to any material, scintillators can measure their
own internal radioactive contamination by themselves. This
provides in general higher sensitivity, and further we will give
more details on methods of analysis of radioactive contaminants in
the data collected with scintillators.

\clearpage
 \begin{landscape}
\begin{table*}[htbp]
\caption{Elements containing primordial radioactive isotopes
(except radium that is originated from $^{238}$U decay). Data on
the half-lives of $\alpha$
 and $\beta$ active isotopes are taken from \cite{ensdf} if other source is not referred;
 year is accepted as 1 yr = 365.2422 d;
 the isotopic abundances are from \cite{Meija:2016a} (the abundance of $^{226}$Ra is assumed to be 100\%);
the atomic weights are from \cite{Meija:2016b}.}
\begin{center}
\resizebox{1.0\textwidth}{!}{
\begin{tabular}{llllll}
 \hline
 Element    & Radioactive                           & Isotopic      & Half-life                                             & Activity in           & Mass of element (g)  \\
 ~          & isotope,                              & abundance     & (yr)                                                  & 1 g of element        & corresponding to 1 mBq \\
 ~          & decay modes                           & ~             &  ~                                                    & (Bq)                  & activity of the \\
 ~          & ~                                     & ~             &  ~                                                    & ~                     & radioactive isotope \\
  \hline
 Potassium  & $^{40}$K, $\varepsilon$, $\beta^-$    & 0.000117(1)   & $1.248(3)\times10^9$                                  & 31.7(3)               & $3.15(3)\times 10^{-5}$  \\
  \hline
 Calcium    & $^{48}$Ca, $2\beta^-$                 & 0.00187(21)   & $6.4^{+1.4}_{-1.1}\times10^{19}$~~\cite{Arnold:2016a}   & $10(2)\times10^{-9}$  & $1.0(3)\times10^{5}$  \\
  \hline
 Vanadium   & $^{50}$V, $\varepsilon$               & 0.00250(10)   & $2.29(25)\times10^{17}$~~\cite{Dombrowski:2011a}       & $2.8(3)\times10^{-6}$ & $352(40)$  \\
  \hline
 Germanium  & $^{76}$Ge, $2\beta^-$                 & 0.0775(12)    & $1.926(94)\times10^{21}$~~\cite{Agostini:2015a}        & $7.1(4)\times10^{-9}$ & $1.41(7)\times 10^5$ \\
  \hline
 Selenium   & $^{82}$Se, $2\beta^-$                 & 0.0882(15)    & $9.6(10)\times10^{19}$~~\cite{Arnold:2005a}            & $1.5(2)\times10^{-7}$& $6.5(7)\times 10^3$ \\
  \hline
 Rubidium   & $^{87}$Rb, $\beta^-$                  & 0.2783(2)     & $4.97(3)\times10^{10}$                                & $867(5)$              & $1.154(7)\times 10^{-6}$ \\
  \hline
 Zirconium  & $^{96}$Zr, $2\beta^-$                 & 0.0280(2)     & $2.35(21)\times10^{19}$~~\cite{Argyriades:2010a}      & $1.73(15)\times10^{-7}$ & $5.8(5)\times 10^{3}$ \\
  \hline
 Molybdenum & $^{100}$Mo, $2\beta^-$                & 0.09744(65)   & $6.9(4)\times10^{18}$~~\cite{Armengaud:2017a}        & $1.95(11)\times10^{-6}$ & $514(31)$ \\
  \hline
 Cadmium    & $^{113}$Cd, $\beta^-$                 & 0.12227(7)    & $8.04(5)\times10^{15}$                                & $1.789(11)\times10^{-3}$ & $0.559(4)$ \\
 ~~          & $^{116}$Cd, $2\beta^-$                & 0.07512(54)   & $2.69(14)\times10^{19}$~~\cite{Polischuk:2017a}        & $3.28(17)\times10^{-7}$ & $3.05(16)\times10^3$ \\
  \hline
 Indium     & $^{115}$In, $\beta^-$                 & 0.95719(52)   & $4.41(25)\times10^{14}$                               & $0.250(14)$           & $4.00(23)\times10^{-3}$ \\
  \hline
 Tellurium  & $^{128}$Te, $2\beta^-$                & 0.3174(8)     & $2.0(3)\times10^{24}$~~\cite{Barabash:2015a}           & $1.65(25)\times10^{-11}$  & $6.1(9)\times10^7$ \\
 ~          & $^{130}$Te, $2\beta^-$                & 0.3408(62)    & $8.2(6)\times10^{20}$~~\cite{Alduino:2017a}            & $4.3(3)\times10^{-8}$ & $2.32(19)\times10^4$ \\
  \hline
 Lanthanum  & $^{138}$La, $\varepsilon$, $\beta^-$  & 0.0008881(71) & $1.03(1)\times10^{11}$                                 & 0.821(11)             & $1.218(15)\times10^{-3}$ \\
  \hline
 Neodymium  & $^{144}$Nd, $\alpha$                  & 0.23798(19)   & $2.29(16)\times10^{15}$                               & $9.5(7)\times10^{-3}$ & 0.105(7) \\
 ~          & $^{150}$Nd, $2\beta^-$                & 0.05638(28)   & $9.34^{+0.66}_{-0.64}\times10^{18}$~\cite{Arnold:2016b}& $5.5(4)\times10^{-7}$& $1.81(13)\times10^3$ \\
  \hline
 Samarium   & $^{147}$Sm, $\alpha$                  & 0.1500(14)    & $1.060(11)\times10^{11}$                              & 124.5(18)               & $8.03(11)\times10^{-6}$ \\
 ~          & $^{148}$Sm, $\alpha$                  & 0.1125(9)     & $6.4^{+1.2}_{-1.3}\times10^{15}$~~\cite{Casali:2016a} & $1.5(3)\times10^{-3}$ & 0.65(13) \\
  \hline
 Europium   & $^{151}$Eu, $\alpha$                  & 0.4781(6)     & $4.6(1.2)\times10^{18}$~~\cite{Casali:2014a}          & $9.0(24)\times10^{-6}$& 111(29) \\
  \hline
 Gadolinium & $^{152}$Gd, $\alpha$                  & 0.0020(3)     & $1.08(8)\times10^{14}$                                & $1.56(27)\times10^{-3}$& 0.64(10)  \\
  \hline
 Lutetium   & $^{176}$Lu, $\beta^-$                 & 0.02599(13)   & $3.640(35)\times10^{10}$~~\cite{Kossert:2013a}         & 54.0(6)               & $1.853(20)\times10^{-5}$ \\
  \hline
 Hafnium    & $^{174}$Hf, $\alpha$                  & 0.0016(12)    & $2.0(4)\times10^{15}$                                 & $6(5)\times10^{-5}$   & 17(13) \\
  \hline
 Tungsten   & $^{180}$W, $\alpha$                   & 0.0012(1)     & $1.8(2)\times10^{18}$~~\cite{Cozzini:2004a}           & $4.8(5)\times10^{-8}$ & $2.08(23)\times10^{4}$ \\
  \hline
 Rhenium    & $^{187}$Re, $\beta^-$                 & 0.6260(5)     & $4.33(7)\times10^{10}$                                & 1027(17)              & $9.74(16)\times10^{-7}$ \\
  \hline
 Osmium     & $^{186}$Os, $\alpha$                  & 0.0159(64)    & $2.0(11)\times10^{15}$                                & $6(4)\times10^{-4}$   & 1.8(12) \\
  \hline
 Platinum   & $^{190}$Pt, $\alpha$                  & 0.00012(2)    & $4.97(16)\times10^{11}$~~\cite{Braun:2017a}            & 0.0164(29)            & 0.061(10) \\
  \hline
 Bismuth    & $^{209}$Bi, $\alpha$                  & 1             & $2.01(8)\times10^{19}$                                & $3.15(13)\times10^{-6}$ & 318(13) \\
  \hline
 Radium     & $^{226}$Ra, $\alpha$                  & 1             & 1600(7)                                               & $3.658(16)\times10^{10}$ & $2.734(12)\times10^{-14}$ \\
  \hline
 Thorium    & $^{230}$Th, $\alpha$                  & 0.0002(2)     & $7.54(3)\times10^{4}$                                 & $1.5(15)\times10^{5}$ & $6.6(66)\times10^{-9}$  \\
 ~          & $^{232}$Th, $\alpha$                  & 0.9998(2)     & $1.40(1)\times10^{10}$                                & 4071(29)              & $2.456(18)\times10^{-7}$  \\
  \hline
 Uranium    & $^{234}$U, $\alpha$                   & 0.000054(5)   & $2.455(6)\times10^{5}$                                & $1.22(12)\times10^{4}$ & $8.2(7)\times10^{-8}$ \\
 ~          & $^{235}$U, $\alpha$                   & 0.007204(6)   & $7.04(1)\times10^{8}$                                 & $568.7(9)$            & $1.7585(29)\times10^{-6}$  \\
 ~          & $^{238}$U, $\alpha$                   & 0.992742(10)  & $4.468(6)\times10^{9}$                                & 12347(17)             & $8.099(11)\times10^{-8}$  \\
 \hline
\end{tabular}
 }
 \end{center}
 \label{tab2}
 \end{table*}
\end{landscape}

\subsection{Indirect methods}
\label{sec:exp-meth-indir}

Long living radioactive isotopes can be measured with the help of
Inductively Coupled Plasma Mass Spectrometry (ICP-MS). Sensitivity
of this method depends on measured matrix and on previous use of
an apparatus. For instance, sensitivity of the spectrometers
installed in the Gran Sasso underground laboratory of I.N.F.N.
(Italy) to the solids and reagents involved in the TeO$_2$
crystals production process was on the level of $\sim
(2\times10^{-10}-2\times10^{-12})$ g/g for $^{232}$Th and
$^{238}$U \cite{Nisi:2017a}, which corresponds to activity of
$^{232}$Th and $^{238}$U: $\sim (0.8-0.008)$ mBq/kg and  $\sim
(2-0.02)$ mBq/kg, respectively. Sensitivity of ICP-MS method to
potassium is hampered due to interference of $^{39}$K with
polyatomic species produced from argon used in the plasma source
of ICP-MS devices \cite{Nisi:2017a}. The measured $^{39}$K
concentrations in NaI powder for radiopure NaI(Tl) crystal
scintillators production were $\sim (1\times10^{-8}-
2\times10^{-7})$ g/g, while the detection limit was on the level
of $<3\times10^{-9}$ g/g~\cite{Nisi:2017a}, that corresponds to
$^{40}$K activity $\sim (0.3-6)$ mBq/kg and a detection limit
$<0.09$ mBq/kg. A higher sensitivity (a detection limit 0.016
mBq/kg of $^{40}$K in NaI matrix) was reported in
\cite{Arnquist:2017a}. The improved sensitivity was achieved
thanks to the use of improved instrumentation, cool plasma
operating conditions, and meticulously clean sample preparations.
Accelerator mass-spectrometry allows to reach much higher
sensitivity up to $\sim10^{-17}$ g/g (potassium in liquid
scintillator \cite{Dong:2007a}). Unfortunately, mass-spectrometry
is practically useless to measure contamination by $^{226}$Ra
(daughter of $^{238}$U) and $^{228}$Th ($^{232}$Th) due to the
rather short half-lives of these isotopes. At the same time,
activities of these radionuclides (those daughters, $^{214}$Bi and
$^{208}$Tl, are the most unfavorable background sources for double
$\beta$ decay experiments) can be substantially different from
$^{238}$U and $^{232}$Th due to broken secular equilibrium. The
same can be said about $^{210}$Pb, daughter of $^{238}$U, one of
the most troublesome contaminant radionuclides for dark matter
experiments.

A very high sensitivity to $^{40}$K, $^{232}$Th and $^{238}$U on
the level of nBq/kg was reached by using neutron-activation
technique to measure contamination of liquid scintillator for the
KamLAND experiment \cite{Djurcic:2003a}. While neutron-activation
analysis is very powerful tool to measure radioactive
contamination in organic scintillators, the method is much less
effective to examine inorganic materials due to possible
activation of scintillation crystals matrix \cite{Heusser:1995a}.
Comparison of the most sensitive indirect methods (mass
spectrometry and neutron activation technique) with direct
counting is discussed in \cite{Povinec:2017a}.

It should be stressed that the various analytical methods used
for analysis of chemical impurities in raw materials for crystal
growth, like for instance Atomic Absorption Spectroscopy, X-ray
Fluorescence, allow to estimate presence of long-living naturally
occurring primordial radioactive elements too. It is worth to
mention also Electron Microscope measurements to estimate presence
of platinum in $^{116}$CdWO$_4$ crystal scintillators
\cite{Danevich:2003b}. However, the sensitivity of these methods
are much below of the ICP-MS sensitivity, not to say for direct
counting methods \cite{Heusser:1995a,Laubenstein:2017a}.

\subsection{Direct detection}
\label{sec:exp-meth-dir}

\subsubsection{Low background measurements}
\label{sec:exp-meth-dir-lbg}

Ultra-low background HP Ge $\gamma$ detectors can be used to
measure radioactive contamination of scintillation crystals and
materials for their production (see, for instance
\cite{Balysh:1993a,Kim:2005a,Belli:2007c,Bavykina:2009a,Barinova:2009a,Loaiza:2011a,Barabash:2011a,Danevich:2011a,Belli:2012b,Belli:2013a,Lee:2007b,Picado:2015a,Cardenas:2017a,Simgen:2017a}).
This method provides typical sensitivity at the level of mBq/kg
for $^{40}$K, $^{137}$Cs, $^{228}$Th, $^{226}$Ra and $^{227}$Ac
(daughter of $^{235}$U), and somewhat lower sensitivity to other
U/Th daughters. However, this approach is useless to detect
internal contamination by $\alpha$ and $\beta$ active
isotopes if decay goes to the ground state of a daughter
nucleus\footnote{Beta activity can be detected by HP Ge detectors
via registration of bremsstrahlung, however the sensitivity in
this case is much lower due to low efficiency and absence of a
clear signature (like peaks in $\gamma$ spectra).}.

The highest sensitivity to measure internal contamination of
scintillators can be achieved in low background measurements where
the scintillator is operating as a detector. Such an approach
provides high detection efficiency, especially for $\alpha$ and
$\beta$ particles. A typical low background scintillation set-up
(see, for instance,
\cite{Danevich:1989a,Ejiri:1991a,Danevich:1996b,Bernabei:1999a,Danevich:2000a,Danevich:2001a,Wang:2002a,Belli:2003a,Ogawa:2003a,Ahmed:2003a,Zhu:2006a,Bernabei:2008c,Polischuk:2017a})
consists of scintillator, light-guide to shield the scintillator
from radioactivity of photomultiplier tubes (typically, the most
contaminated details of a low background scintillation set-up),
passive shield. Background of the detector can be further
suppressed by using of active shield counters surrounding the main
detector, and anti-muon veto
counters\cite{Danevich:2003a,Barabash:2011a}. Light-guides made of
a scintillation material with different (relative to the main
scintillation detector) scintillation decay time can serve as
active anticoincidence detectors
\cite{Danevich:2003a,Barabash:2011a,Belli:2016a}. Continuous
flushing of internal volume of a set-up by a radon-free gas
(typically by nitrogen) allows to reduce background caused by
radon \cite{Belli:2003a,Bernabei:2008c}. It is worth mentioning a
possibility to reduce background further by using data on time of
arrival and pulse-shape of scintillation signals (the methods will
be considered in Sections \ref{sec:exp-meth-dir-t-A} and
\ref{sec:exp-meth-dir-PSD}).

Below we will discuss response of a scintillation detector to
$\gamma$ quanta, $\beta$ and $\alpha$ particles, pulse-shape
discrimination, time-amplitude analysis, Monte-Carlo simulation of
background components. These methods allow to describe background
energy spectra accumulated with a scintillation counter and
estimate radioactive contamination of the scintillator.

\subsubsection{Response of scintillation detector to $\gamma$ quanta, $\beta$ and $\alpha$ particles}
\label{sec:exp-meth-dir-resp}

Knowledge of a detector response to $\gamma$ quanta (response
function, dependence of energy resolution on energy) and $\alpha$
particles (energy dependence of  $\alpha/\beta$
ratio\footnote{Here we define the ``$\alpha/\beta$  ratio'' as the
ratio of $\alpha$ peak position measured in the $\gamma$ energy
scale to the energy of $\alpha$ particles. As usual, a detector
energy scale is measured with $\gamma$ sources, thus the notation
``$\alpha/\gamma$ ratio'' is more adequate. However, because
$\gamma$ rays interact with matter by means of the energy transfer
to electrons, in the present paper we are using a more traditional
term ``$\alpha/\beta$ ratio''.} and energy resolution) is
necessary to interpret background of the detector.

\begin{figure}
\begin{center}
\resizebox{0.55\textwidth}{!}{\includegraphics{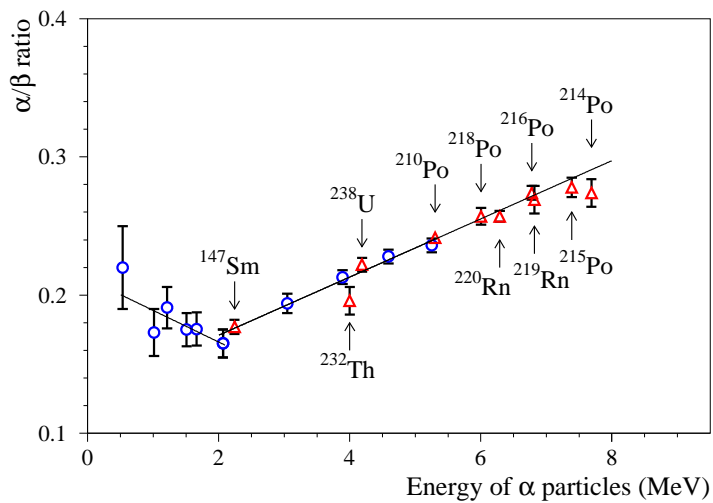}}
 \caption{The energy dependence of the $\alpha/\beta$ ratio measured with CaWO$_4$ crystal scintillator \cite{Zdesenko:2005a}.
 The points obtained by irradiation of the scintillator by external $\alpha$ particles are shown by circles, while the points
 obtained by analysis of $\alpha$ peaks from the internal contamination of the crystal by $\alpha$ active nuclides are drawn by triangles.}
 \label{fig:fig01}
\end{center}
\end{figure}

Response function and dependence of energy resolution on energy of
$\gamma$ quanta can be measured with $\gamma$ sources in a wide
energy interval from a few keV (5.9 keV Mn K X-rays from
$^{55}$Fe) up to 2615 keV ($\gamma$ quanta of $^{208}$Tl).
Calibration with $\alpha$ sources is much more complicated task
because energies of commonly used $\alpha$ sources lie in the
energy region from $\sim5.3$ to $\sim8.8$ MeV ($^{228}$Th,
$^{241}$Am, $^{244}$Cm, $^{252}$Cf). To calibrate detector at
lower energies, $\alpha$ sources with absorbers can be used (see,
for instance
\cite{Danevich:2003b,Belli:2003a,Zdesenko:2005a,Danevich:2005a,Belli:2007b}).
Response of scintillation detectors to $\alpha$ particles is
non-linear. An example of $\alpha/\beta$ ratio dependence on energy is
shown in Fig. \ref{fig:fig01}, where the $\alpha/\beta$ ratio
measured with a CaWO$_4$ crystal scintillator is presented. As the
quenching of the scintillation light caused by $\alpha$ particles is due to
the higher ionization density, such a
behavior of the $\alpha/\beta$ ratio can be explained by the
energy dependence of ionization density of particles
\cite{Birks,Tretyak:2010a}.
See recent review \cite{Wolszczak:2017} on $\alpha/\beta$ ratio in different
scintillators.

\nopagebreak

\begin{figure}
\begin{center}
\resizebox{0.55\textwidth}{!}{\includegraphics{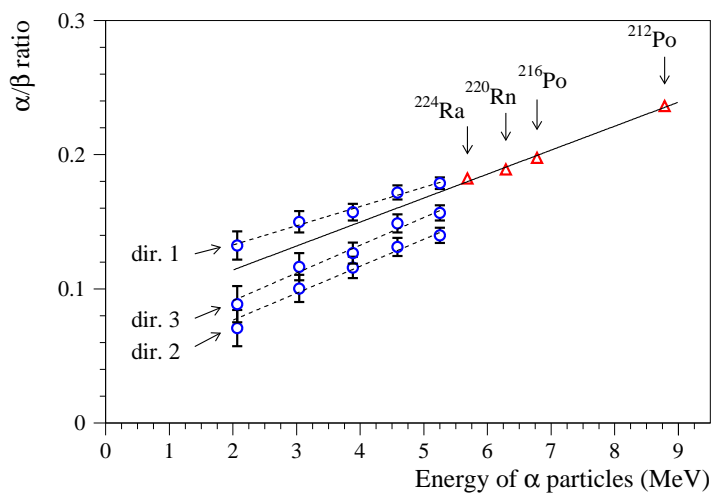}}
\caption{Energy dependence of the $\alpha/\beta$ ratio on energy
measured with $^{116}$CdWO$_4$ crystal scintillator. The
$\alpha/\beta$ ratio depends on direction of irradiation
relatively to crystal axes denoted as dir. 1 (along to the crystal
axis [010]), dir. 2 ([001]) and dir. 3 ([100])
\cite{Danevich:2003b}. The points obtained by irradiation of the
scintillator by external $\alpha$ particles in the directions
parallel to the axes of the crystal are shown by circles, while
the points obtained by analysis of $\alpha$ peaks from internal
contamination of the scintillator by $\alpha$ active nuclides are
drawn by triangles.}
 \label{fig:fig02}
\end{center}
\end{figure}

Alpha peaks from internal contamination of scintillators allow to
extend the interval of $\alpha$ particles energies (see Fig.
\ref{fig:fig01} and Fig. \ref{fig:fig02}). In addition, $\alpha$
peaks from internal contamination provide an important test of
calibration measurements with external sources. It should be
stressed that analysis of internal $\alpha$ peaks is only
practical method to measure response to $\alpha$ particles for
encapsulated scintillation detectors produced from highly
hygroscopic materials, like NaI(Tl), LaCl$_3$(Ce), LaBr$_3$, etc.

\nopagebreak

In some scintillation crystals with anisotropic structure,
$\alpha/\beta$ ratio depends on the direction of $\alpha$
particles relatively to crystal axes. Such an effect was observed
in anthracene \cite{Schuster:2016}, stilbene \cite{Schuster:2017}, and
CdWO$_4$ \cite{Danevich:2003b}, ZnWO$_4$ \cite{Danevich:2005a},
and MgWO$_4$ \cite{Danevich:2009a} crystal scintillators (see Fig.
\ref{fig:fig02} where dependence of $\alpha/\beta$ ratio on
direction of $\alpha$ irradiation relatively to crystal axes of
CdWO$_4$ scintillator is presented). It leads to some worsening of
energy resolution of these detectors to $\alpha$ particles
\cite{Danevich:2003b,Belli:2011a}.

Scintillation signals for $\gamma$ quanta can be quenched in
comparison to $\beta$ particles. E.g. there is an indication of
scintillation light-efficiency quenching in CdWO$_4$ scintillator
for $\gamma$ quanta in comparison to electrons that resulted in a
higher $Q_{\beta}$ value of $^{113}$Cd
\cite{Danevich:1996a,Belli:2007a}: $Q_{\beta}=337$ keV and
$Q_{\beta}=345$ keV, respectively. These values are substantially
larger than the Table value $Q_{\beta}=323.83(27)$ keV
\cite{Wang:2017a}. The quenching can be explained by the
non-proportionality in the scintillation response of CdWO$_4$
crystal scintillators significant for energies below $\sim 0.1$
MeV \cite{Bizzeti:2012a}. Because of distribution of $\gamma$
quanta energy between two and more electrons (due to the multiple
Compton scattering), $\gamma$ peaks should be shifted to lower
energies in ``electron'' energy scale\footnote{This effect also
leads to worse energy resolution of scintillation detectors for
$\gamma$ quanta \cite{Dorenbos:1995a,Moszynski:2016a}.}. Organic
scintillators, particularly liquid scintillators, have significant
non-proportionality of scintillation response. The effect was
observed in liquid scintillators used for neutrino experiments
Borexino \cite{Back:2002a} and Double Chooz \cite{Aberle:2011a}.
The effect of $\gamma$ peaks shift in liquid scintillators is
significant: e.g. the position of the 1461 keV $\gamma$ peak of
$^{40}$K  in the scintillator used in the Borexino detector is
shifted to 1360 keV relatively to the energy deposited for
electrons \cite{Back:2003a}.

\subsubsection{Time-amplitude analysis}
\label{sec:exp-meth-dir-t-A}

Data on the energy and arrival time of events can be used to
select fast decay chains from the $^{232}$Th, $^{235}$U and
$^{238}$U families. The method of time-amplitude analysis is
described in detail in
\cite{Danevich:1995a,Barton:2000a,Danevich:2001a,Danevich:2003a,Zhu:2006a}.
For instance, the following sequence of $\alpha$ decays from the
$^{232}$Th family can be selected:

\begin{center}
$^{224}$Ra ($Q_{\alpha}=5789$ keV; $T_{1/2} = 3.632$ d)
$\rightarrow$ $^{220}$Rn ($Q_{\alpha} = 6405$ keV; $T_{1/2} =
55.6$ s) $\rightarrow$ $^{216}$Po ($Q_{\alpha} = 6906$ keV;
$T_{1/2} = 0.145$ s) $\rightarrow$ $^{212}$Pb.
\end{center}

\begin{figure}
\begin{center}
\resizebox{0.6\textwidth}{!}{\includegraphics{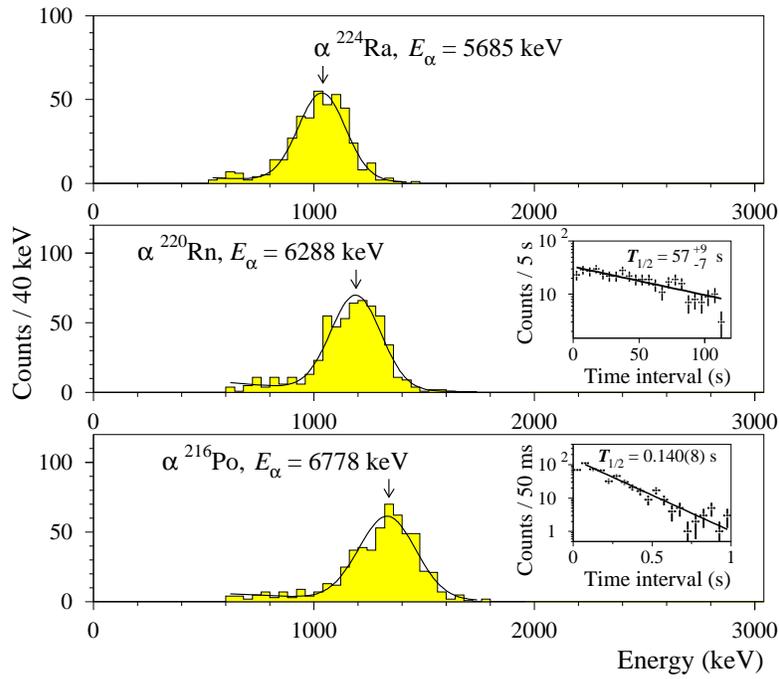}}
\caption{The $\alpha$ peaks of $^{224}$Ra, $^{220}$Rn, and
$^{216}$Po selected by the time-amplitude analysis from the data
accumulated during 14745 h with $^{116}$CdWO$_4$ detector
\cite{Danevich:2003b}. (Insets) The time distributions between the
first and second (and between second and third) events are
presented together with exponential fits. Obtained half-lives of
$^{220}$Rn and $^{216}$Po ($57^{+9}_{-7}$ s and $0.140(8)$ s,
respectively) are in a good agreement with the table values
(55.6(1) s and 0.145(2) s, respectively \cite{ensdf}).}
 \label{fig:fig03}
\end{center}
\end{figure}

These radionuclides are in equilibrium with $^{228}$Th ($^{232}$Th
family). As an example, the results of the time-amplitude analysis
of data accumulated in the low background experiment to search for
$2\beta$ decay of $^{116}$Cd with the help of $^{116}$CdWO$_4$
crystal scintillators are shown in Fig. \ref{fig:fig03}. The
obtained $\alpha$ peaks (the $\alpha$ nature of events was
confirmed by the pulse-shape analysis described in Section
\ref{sec:exp-meth-dir-PSD}) as well as the distributions of the
time intervals between events, are in a good agreement with those
expected for the $\alpha$ decays of the $^{224}$Ra $\rightarrow$
$^{220}$Rn $\rightarrow$ $^{216}$Po $\rightarrow$ $^{212}$Pb
chain.

\nopagebreak

Similarly the following fast sequence of $\beta$ and $\alpha$
decays:

\begin{center}
$^{214}$Bi ($Q_{\beta}=3269$ keV; $T_{1/2} = 19.9$ m)
$\rightarrow$ $^{214}$Po ($Q_{\alpha} = 7834$ keV; $T_{1/2} =
164.3$ $\mu$s) $\rightarrow$ $^{210}$Pb
\end{center}

\begin{figure}
\begin{center}
\resizebox{0.6\textwidth}{!}{\includegraphics{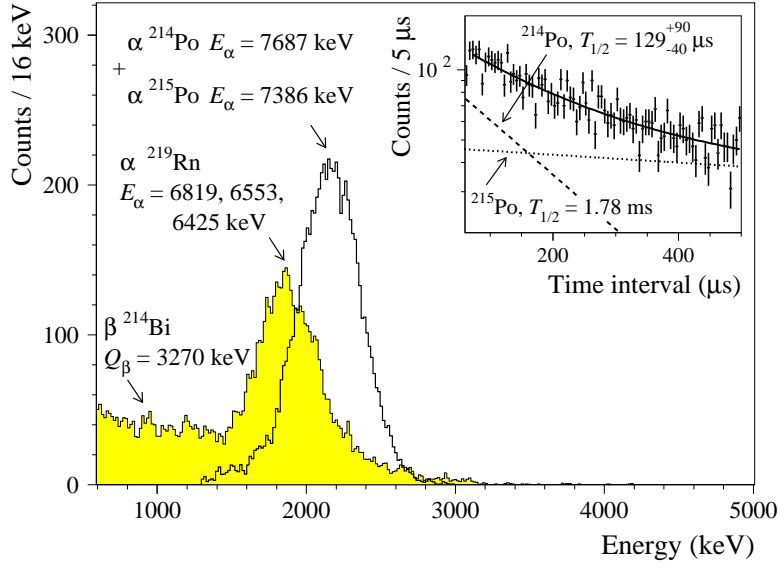}}
\caption{The energy spectra of the sequence of $\beta$ and
$\alpha$ decays in the decay chain $^{214}$Bi $\rightarrow$
$^{214}$Po $\rightarrow$ $^{210}$Pb ($^{238}$U family) which were
selected with the help of the time-amplitude analysis of 8609 h
data accumulated with GSO scintillator \cite{Danevich:2001a}. The
peak with the energy in $\gamma$ scale $\approx1.8$ MeV is related
with the decays of $^{219}$Rn from the chain $^{219}$Rn
$\rightarrow$ $^{215}$Po $\rightarrow$ $^{211}$Pb ($^{235}$U
family). In the insert: the distribution of the time intervals
between the first and second events together with its fit (solid
line) by the sum of exponent with $T_{1/2}=129^{+90}_{-40}$ $\mu$s
(a table $^{214}$Po value is $T_{1/2}=164.3$ $\mu$s; dashed line)
and exponent with $T_{1/2}=1.78$ ms corresponding to decays of
$^{215}$Po from the chain $^{219}$Rn $\rightarrow$ $^{215}$Po
$\rightarrow$ $^{211}$Pb (dotted line).}
 \label{fig:fig04}
\end{center}
\end{figure}

\noindent (in equilibrium with $^{226}$Ra from the $^{238}$U
family) can also be selected with the help of time-amplitude
analysis. In Fig. \ref{fig:fig04} one can see the energy spectra
and time distributions of the sequence selected from the data
accumulated in the low background experiment to search for
$2\beta$ decay of $^{160}$Gd with the help of gadolinium
orthosilicate (Gd$_2$SiO$_5$:Ce, GSO) scintillator
\cite{Danevich:2001a}. In addition, Fig. \ref{fig:fig04}
illustrates a possibility to detect another short chain:

\begin{center}
$^{219}$Rn ($Q_{\alpha}=6946$ keV; $T_{1/2} = 3.96$ s)
$\rightarrow$ $^{215}$Po ($Q_{\alpha} = 7526$ keV; $T_{1/2} =
1.781$ ms) $\rightarrow$ $^{211}$Pb.
\end{center}

\noindent Radionuclide $^{219}$Rn is in equilibrium with
$^{227}$Ac from the $^{235}$U family. In this case the events of
$^{214}$Po and $^{215}$Po decays are superimposed (see Fig.
\ref{fig:fig04}). Nevertheless activities of $^{226}$Ra and
$^{227}$Ac can be calculated separately thanks to possibility to
distinguish between the broad $\beta$ spectrum of $^{214}$Bi and
$\alpha$ peak of $^{219}$Rn.

%\clearpage
\nopagebreak

\subsubsection{Pulse-shape discrimination}
\label{sec:exp-meth-dir-PSD}

Most of scintillators have slightly different decay kinetic for
$\beta$ particles ($\gamma$ quanta) and $\alpha$ particles. It
allows to discriminate these particles, and therefore, to estimate
activity of $\alpha$ and $\beta$ active nuclides separately.
Different methods can be
used to realize pulse-shape discrimination. We would like to refer
to the optimal filter method proposed by Gatti and De Martini
\cite{Gatti}, and developed in \cite{Fazzini:1998a} for CdWO$_4$
crystal scintillators, and then successfully applied to many
scintillators
\cite{Belli:2003a,Annenkov:2008a,Danevich:2005a,Danevich:2003b,Zdesenko:2005a,Belli:2007b,Danevich:2005b,Bardelli:2008a,Danevich:2009a}.
In the optimal filter method, a shape indicator ($SI$) - a
numerical characteristic - can be calculated for each
scintillation event produced by a scintillator:

\begin{equation}
SI=\sum[f(t_k)\times P(t_k)]/\sum f(t_k),
 \label{eq:1}
\end{equation}

\noindent where the sum is over time channels $k$ from the origin
of the pulse up to a certain time; $f(t_k)$ is the digitized
amplitude of the signal (at the time $t_k$). The weight function
$P(t_k)$ is defined as:

\begin{equation}
P(t_k)=[f_{\alpha}(t_k)-f_{\gamma}(t_k)]/[f_{\alpha}(t_k)+f_{\gamma}(t_k)],
 \label{eq:2}
\end{equation}

\noindent where $f_{\alpha}(t_k)$ and $f_{\gamma}(t_k)$ are the
reference pulse shapes for $\alpha$ particles and $\gamma$ quanta.
To realize the optimal filter method, the pulse shapes for
$\alpha$ particles and $\gamma$ quanta should be studied. The
distributions of the shape indicators have Gaussian shape.
Therefore, the figure of merit ($FOM$) -- a measure of
discrimination ability -- can be calculated using the following
expression proposed by Gatti and De Martini:

\begin{equation}
FOM=|SI_{\alpha}-SI_{\gamma}|/\sqrt{\sigma_{\alpha}^2+\sigma_{\gamma}^2}.
 \label{eq:3}
\end{equation}

\begin{figure}
\begin{center}
\resizebox{0.55\textwidth}{!}{\includegraphics{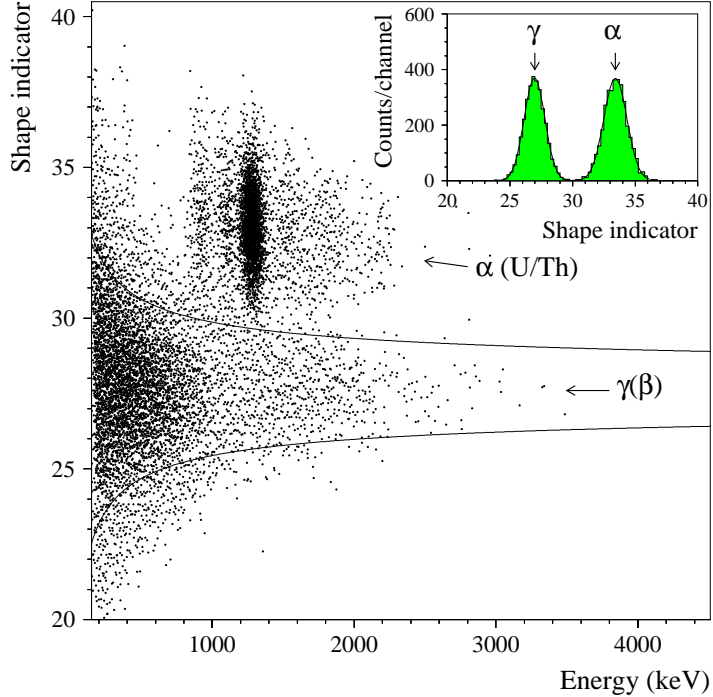}}
\caption{Scatter plot of the shape indicator $SI$ (see text)
versus energy for 171 h of the background data measured with
CaWO$_4$ crystal scintillator. Lines show $\pm2\sigma$ region of
$SI$ for $\gamma(\beta)$ events. (Inset) The $SI$ distributions
measured in calibration runs with $\alpha$ particles
($E_{\alpha}=5.3$ MeV) and $\gamma$ quanta ($\approx1.2$ MeV)
\cite{Zdesenko:2005b}.}
 \label{fig:fig05}
\end{center}
\end{figure}

\begin{figure}
\begin{center}
\resizebox{0.6\textwidth}{!}{\includegraphics{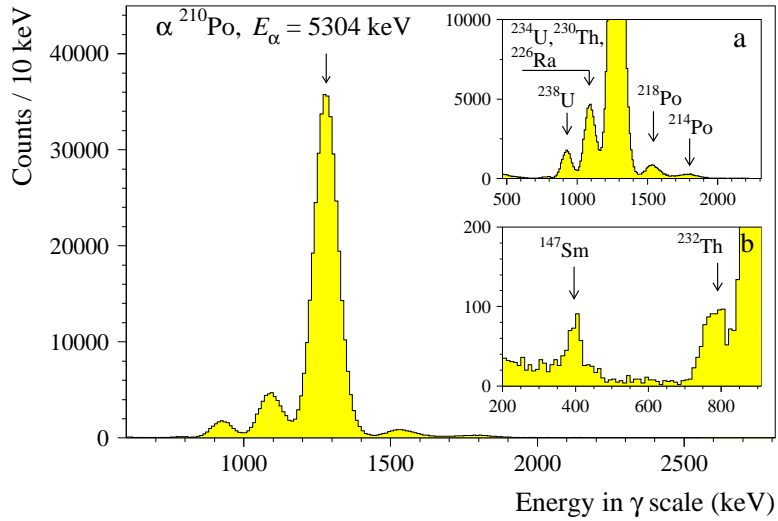}}
\caption{Energy spectrum of $\alpha$ events selected by the
pulse-shape analysis from background data measured over 1734 h
with CaWO$_4$ detector \cite{Zdesenko:2005b}. (Inset a) The same
spectrum but scaled up. It is well reproduced by the model, which
includes $\alpha$ decays of nuclides from $^{232}$Th and $^{238}$U
families. (Inset b) Low energy part of the spectrum where an
$\alpha$ peak of $^{147}$Sm is clearly visible.}
 \label{fig:fig06}
\end{center}
\end{figure}

\begin{figure}
\begin{center}
\resizebox{0.6\textwidth}{!}{\includegraphics{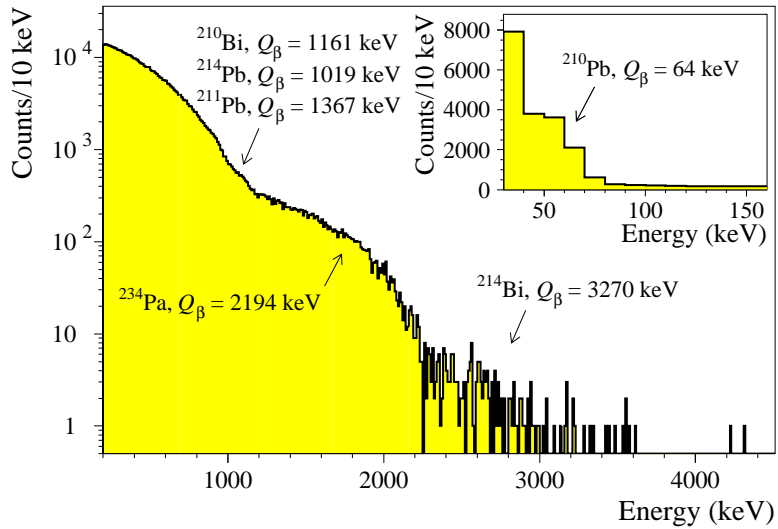}}
\caption{Energy spectrum of $\beta(\gamma)$ events selected by the
pulse-shape analysis technique from the background data measured
over 1734 h with CaWO$_4$ detector \cite{Zdesenko:2005b}. The
distribution is described by the $\beta$ spectra of $^{210}$Bi,
$^{214}$Pb, $^{211}$Pb, $^{234m}$Pa, and $^{214}$Bi. (Inset) In
the low energy region, the background (measured over 15.8 h) is
caused mainly by decay of $^{210}$Pb.}
 \label{fig:fig07}
\end{center}
\end{figure}

One can see an illustration of pulse-shape discrimination by using
the optimal filter method in Fig. \ref{fig:fig05} where the
scatter plot of the shape indicator versus energy is shown for the
data measured with a CaWO$_4$ crystal scintillator
\cite{Zdesenko:2005b}. Energy spectrum of $\alpha$ events selected
from the data measured with the CaWO$_4$ crystal over 1734 h is
presented in Fig. \ref{fig:fig06}, while an energy spectrum of
$\beta(\gamma)$ events is shown in Fig. \ref{fig:fig07}.

\nopagebreak

\nopagebreak

\nopagebreak

The mean time method is also widely used to discriminate
$\beta(\gamma)$ and $\alpha$ particles in scintillation detectors
(see e.g.
\cite{Lee:2007b,Bernabei:1999a,Zhu:2006a,Bardelli:2008a}). The
following formula can be applied to calculate the mean time
parameter $\langle t \rangle$ for each pulse:

\begin{equation}
 \langle t \rangle = \sum (f(t_k)\times t_k)/\sum f(t_k),
 \label{eq:4}
\end{equation}

\noindent where the sum is over time channels $k$, starting from
the origin of pulse and up to a certain time. The distributions of
parameters $\langle t \rangle$ for $\alpha$ and $\beta(\gamma)$
signals are also well described by Gaussian functions. Thus, the
same measure ($FOM$, see Equation \ref{eq:3}) can be used to
estimate efficiency of the method. The optimal filter method
provides slightly better pulse-shape discrimination in comparison
with the mean time technique (see e.g. \cite{Bardelli:2008a}).
However, the mean time method is easier to apply because it does
not require the construction of a weight function, which requires
knowledge of scintillation signals pulse-shape.

It should be stressed that the pulse-shape of scintillation
signals for $\alpha$ particles depends on energy. In some
scintillators with anisotropic properties pulse-shape also depends
on direction of $\alpha$ irradiation relatively to the crystal
axes. As in the case with the $\alpha/\beta$ ratio, such a dependence
was observed in CdWO$_4$ \cite{Danevich:2003b}, ZnWO$_4$
\cite{Danevich:2005a}, and MgWO$_4$ \cite{Danevich:2009a} crystal
scintillators.

Pulse-shape analysis can also be applied to the very fast sequence
of decays from the $^{232}$Th family:

\begin{center}
$^{212}$Bi ($Q_{\beta}=2252$ keV; $T_{1/2} = 60.55$ m)
$\rightarrow$ $^{212}$Po ($Q_{\alpha} = 8954$ keV; $T_{1/2} =
0.299$ $\mu$s) $\rightarrow$ $^{208}$Pb.
\end{center}

\begin{figure}
\begin{center}
\resizebox{0.55\textwidth}{!}{\includegraphics{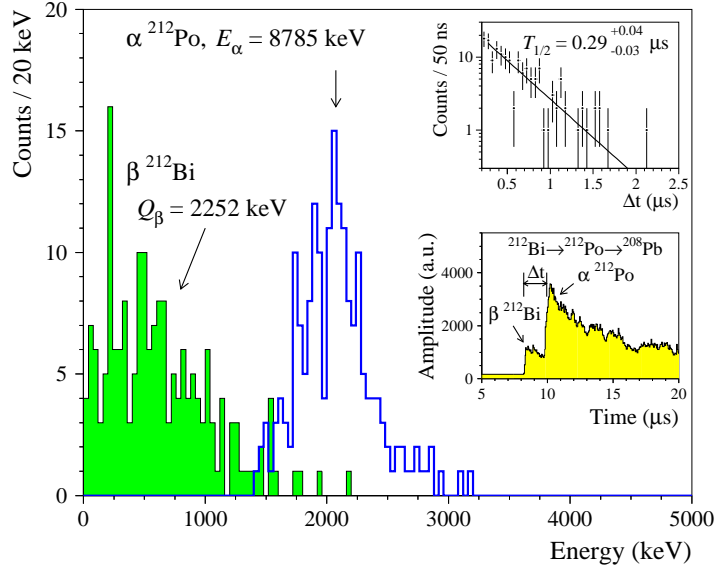}}
\caption{The energy distributions for the fast sequence of $\beta$
($^{212}$Bi, $Q_{\beta}=2252$ keV) and $\alpha$ ($^{212}$Po,
$E_{\alpha}=8785$ keV, $T_{1/2}=0.299$ s) decays selected by the
pulse-shape analysis from the background data obtained in the
experiment with cadmium tungstate crystal scintillators
\cite{Danevich:2003b}. (Upper inset) The time distribution of
intervals ($\Delta t$) between $\beta$ and $\alpha$ signals.
(Lower inset) Example of such an event in the CdWO$_4$
scintillator.}
 \label{fig:fig08}
\end{center}
\end{figure}

An example of such an analysis is presented in Fig.
\ref{fig:fig08}, where the $\beta$ spectrum of $^{212}$Bi, the
$\alpha$ peak of $^{212}$Po and the distribution of the time
intervals between the first and the second pulses selected from the
data of low background experiment \cite{Danevich:2003b} are
depicted. This method is more effective with fast scintillators.

\nopagebreak

\subsubsection{Simulation of background energy spectra}
\label{sec:exp-meth-dir-MCS}

\begin{figure}
\begin{center}
\resizebox{0.6\textwidth}{!}{\includegraphics{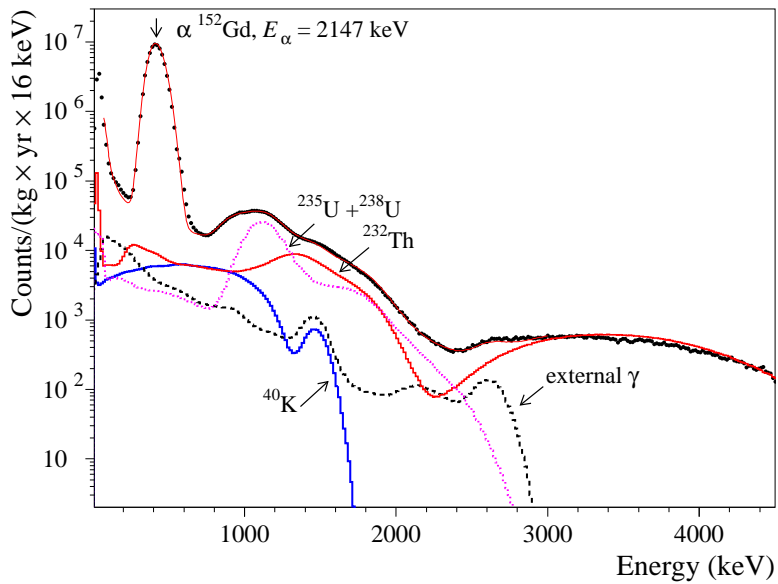}}
\caption{The background spectrum of the GSO detector for 0.969 yr
kg of exposure (points) and the model of background (solid line)
obtained by the fit of the data in the $60-3000$ keV energy
interval \cite{Danevich:2001a}. The most important internal
($^{40}$K, sum of $^{235}$U and $^{238}$U, $^{232}$Th) and
external ($\gamma$ radiation from the photomultiplier tube)
components of background are shown. A peak at the energy
$\approx420$ keV is due to the $\alpha$ activity of $^{152}$Gd,
while the background in the energy interval $0.8-2$ MeV is caused
mainly by $\alpha$ decays of U/Th daughters.}
 \label{fig:fig09}
\end{center}
\end{figure}

To estimate possible contamination of a scintillator, especially
by $\beta$ active nuclides, one can fit an energy spectrum by
Monte Carlo simulated models. As an example, the fit of the low
background energy spectrum accumulated with GSO crystal
scintillator in the experiment to search for $2\beta$ decay of
$^{160}$Gd \cite{Danevich:2001a} is presented in Fig.
\ref{fig:fig09}. The models of background were simulated with the
help of the GEANT package \cite{Agostinelli:2003a,Allison:2006a,Allison:2016}.
For models presented in Fig.
\ref{fig:fig09}, an event generator DECAY0 \cite{DECAY0} was used.
This generator allows to take into account a number and types of
emitted particles, their energies, directions of movement and
times of emission. GEANT and EGS4 \cite{EGS4} packages are the
most widely used to simulate background of scintillation
detectors.

\nopagebreak

\subsubsection{Low-temperature scintillating bolometers}
\label{sec:low-temp}

An extremely high sensitivity to $\alpha$ radioactive
contamination of crystal scintillators can be reached by operation
of a scintillator as low temperature scintillating bolometer
thanks to a very high energy resolution and particle
discrimination capability
\cite{Alessandrello:1998a,Pirro:2006a,Gorla:2008a,Poda:2017b}. The
discrimination can be realized by analysis of heat and
scintillation signals from a crystal scintillator (by using a
different scintillation yield for ions \cite{Tretyak:2010a}, in
particular for $\alpha$ particles) or by only thermal pulse
profile analysis \cite{Gironi:2010a}.

\begin{figure}
\begin{center}
\resizebox{0.55\textwidth}{!}{\includegraphics{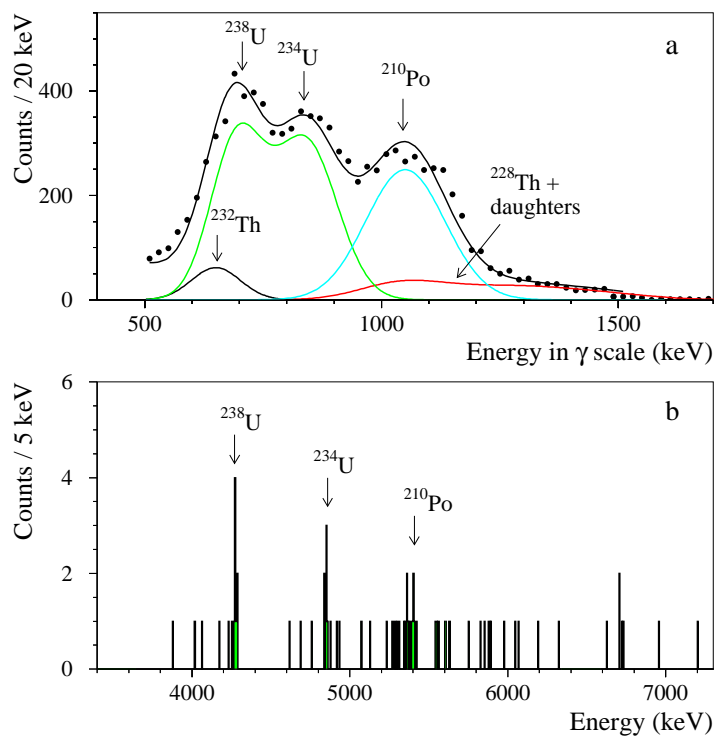}}
\caption{(a) Energy spectrum of $\alpha$ events selected by
pulse-shape discrimination from the data acquired by 589 g
$^{116}$CdWO$_4$ scintillation detector at the Gran Sasso
underground laboratory over 1727 h together with the model which
includes $\alpha$ active nuclides of $^{238}$U and $^{232}$Th
families \cite{Barabash:2011a}. (b) Energy spectrum of $\alpha$
events measured by 34.5 g $^{116}$CdWO$_4$ scintillating bolometer
in a surface laboratory over 250 h \cite{Barabash:2016a}. An
advantage of the high energy resolution and particle
discrimination capability of the low-temperature scintillating
bolometer is clearly visible.}
 \label{fig:fig10}
\end{center}
\end{figure}

The capability of low-temperature scintillating bolometers to
screen $\alpha$ radioactive contamination of crystal scintillators
is illustrated in Fig. \ref{fig:fig10}, where $\alpha$ spectra
gathered with two samples of $^{116}$CdWO$_4$ scintillator
operated as a conventional scintillation detector and as a
cryogenic scintillating bolometer are presented. A small sample of
$^{116}$CdWO$_4$ crystal (34.5 g) was measured over 250 h in a
surface set-up \cite{Barabash:2016a}. The achieved sensitivity on
the level of $\sim0.1$ mBq/kg for $^{232}$Th, $^{235}$U and
$^{238}$U (and their $\alpha$ active daughters) competes with
results of the long time measurements (1727 h) with much larger
mass $^{116}$CdWO$_4$ crystal scintillator (589 g) by using
ultra-low background set-up deep underground at the Gran Sasso
underground laboratory \cite{Barabash:2011a}.

\nopagebreak

\section{Data on radioactive contamination of scintillators}
\label{sec:data}

\begin{figure}
\begin{center}
\resizebox{0.55\textwidth}{!}{\includegraphics{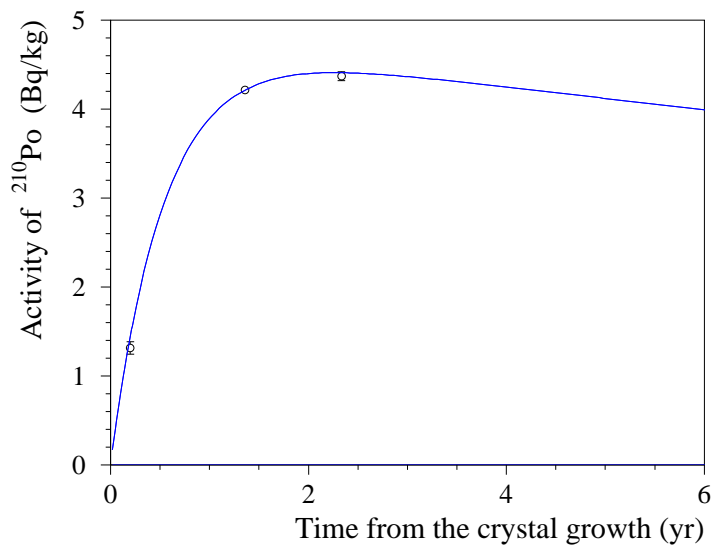}}
\caption{Time dependence of $^{210}$Po activity in CaWO$_4$
crystal scintillator \cite{Danevich:2011a} (circles). The data is
approximated by a function, describing the increase
in $^{210}$Po due to the decay of $^{210}$Pb assuming essentially
lower amount of $^{210}$Po after the crystal growth (solid
line). The $^{210}$Po activity is beginning to decline after
$\sim2.2$ yr due to decay of $^{210}$Pb ($T_{1/2}=22.2$ yr).}
 \label{fig:fig11}
\end{center}
\end{figure}

Data on radioactive contamination of different scintillators are
presented in Tables \ref{tab3}-\ref{tab13}. The main sources of
radioactive contamination of scintillation materials are naturally
occurring radionuclides of the $^{232}$Th and $^{238}$U families,
and $^{40}$K. Activity of $^{235}$U with daughters is observed in
some materials too. It should be stressed that the secular
equilibrium of the U/Th chains is usually broken in scintillation
materials. It means that the activities of $^{238}$U, $^{230}$Th,
$^{226}$Ra, $^{210}$Pb in the $^{238}$U family should be
considered separately. The activity of $^{210}$Po should be also
reported separately if a scintillation material was produced
recently in comparison to the half-life of $^{210}$Po
($T_{1/2}=138.376(2)$ d). A time behaviour of $^{210}$Po activity
in a CaWO$_4$ crystal scintillator, in the case of essentially
lower amount of $^{210}$Po after the crystal growth, is shown in
Fig. \ref{fig:fig11}. Similarly, contaminations by $^{232}$Th,
$^{228}$Ra and $^{228}$Th from the $^{232}$Th family, and
activities of $^{235}$U, $^{231}$Pa and $^{227}$Ac from the
$^{235}$U family can be different. As an example of a strong
disequilibrium of the $^{232}$Th chain we can refer the CeF$_3$
crystal scintillator where the activity of $^{232}$Th exceeded the
activity of $^{228}$Th by a factor 18 \cite{Belli:2003a}. As for
the $^{238}$U chain, its equilibrium is strongly broken e.g. in
CaWO$_4$ \cite{Zdesenko:2005a} where the activities of $^{238}$U,
$^{226}$Ra and $^{210}$Po relate as $1:0.4:21$. Both the
$^{232}$Th and $^{238}$U chains were found strongly broken in
BaF$_2$ crystal scintillator, where radium accumulation
($^{226}$Ra from the $^{238}$U chain, and $^{228}$Ra from the
$^{232}$Th chain) in the crystal was observed \cite{Belli:2014a}.
For this reason a reference date should be given for activities of
relatively short living isotopes $^{210}$Po, $^{228}$Th,
$^{228}$Ac.

\nopagebreak

\clearpage
 \begin{landscape}
\begin{table*}[htbp]
\caption{Radioactive contamination of organic scintillators.}
\begin{center}
\resizebox{1.1\textwidth}{!}{
\begin{tabular}{|l|l|l|l|l|l|l|l|}
 \hline
 Chain      & Source        & \multicolumn{6}{|c|} {Activity (mBq/kg) in sample, Reference}   \\
 \cline{3-8}
 ~          & ~             &  \multicolumn{3}{|c|} {Liquid scintillators}                                      & \multicolumn{3}{|c|} {Plastic scintillators} \\
 \cline{3-8}
 ~          & ~             & \multicolumn{2}{|c|} {Borexino, 278 t}            & KamLAND \cite{Keefer:2011a}   & Polystyrene-based scintillator    & EJ-200                & NE-102A  \\
 \cline{3-4}
 ~          & ~             & Phase I$^a$               & Phase II              &                               & \cite{Argyriades:2010b}          & \cite{Heusser:2015a}  & \cite{Heusser:2015a}\\
 ~          & ~             &\cite{Zuzel:2015a}         & \cite{Zuzel:2015a}    & ~                             &                                   &                       &  \\
 \hline
  ~         & $^{14}$C      & $0.445(10)$               & $0.445(10)$           & 0.66(16)                      &                                   &                       & \\
\hline
  ~         & $^{39}$Ar     & ~                         &                       & $<0.005$                      &                                   &                       & \\
 \hline
  ~         & $^{40}$K      & $<1.1\times10^{-7}$       & $<1.1\times10^{-7}$   & $3.4(3)\times10^{-5}$         &                                   & 1.5(5)                & 2.2(7) \\
\end{tabular}
 }
 \end{center}
 \label{tab3}
 \end{table*}
\end{landscape}

\clearpage
 \begin{landscape}
\begin{table*}[htbp]
\caption{Radioactive contamination of fluorite crystal
scintillators.}
\begin{center}
\resizebox{1.1\textwidth}{!}{
\begin{tabular}{|l|l|l|l|l|l|l|l|l|}
 \hline
 Chain      & Source        & \multicolumn{7}{|c|} {Activity (mBq/kg) in sample, Reference}   \\
 \cline{3-9}
 ~          & ~             & LiF(W)                    & \multicolumn{2}{|c|} {CaF$_2$(Eu)}                & SrF$_2$                   & CeF$_3$                   & \multicolumn{2}{|c|} {BaF$_2$} \\
 \cline{3-9}
 ~          & ~             & 553 g \cite{Belli:2012c}  & 370 g \cite{Belli:2007b}  & 0.29 kg, 25 samples   & 53 g \cite{Diaz:2013a}    & 49 g \cite{Belli:2003a}   & 734 g \cite{Cerulli:2004} & 1.714 kg \cite{Belli:2014a} \\
 ~          & ~             & ~                         & ~                         & \cite{Ogawa:2003a,Ogawa:2004a} & ~                & ~                         & ~                         &  \\
 \hline
  ~         & $^{40}$K      & $<5.1$                    & $<7$                      &                       &                           & $<330$                    &                           & \\
 \hline
  ~         & $^{90}$Sr     &                           & $<4$                      &                       &                           &                           &                           & \\
 \hline
  ~         & $^{137}$Cs    & $<0.2$                    & $<0.3$                    &                       &                           &                           &                           & \\
\hline
  ~         & $^{138}$La    &                           &                           &                       &                           & $<60$                     &                           & \\
\hline
\end{tabular}
 }
 \end{center}
 \label{tab4}
 \end{table*}
\end{landscape}

\begin{table}[ht]
\caption{Radioactive contamination of Y$_3$Al$_5$O$_{12}$:Nd
[YAG(Nd)] and Li$_6$Eu(BO$_3$)$_3$ crystal scintillators.}
\begin{center}
\begin{tabular}{|l|l|l|l|l|}
 \hline
 Chain      & Source        & \multicolumn{3}{|c|} {Activity (mBq/kg) in sample, Reference}   \\
 \cline{3-5}
 ~          & ~             &  YAG(Nd)              & \multicolumn{2}{|c|}{Li$_6$Eu(BO$_3$)$_3$}  \\
 \cline{3-5}
 ~          & ~             &  7.16 g               & 2.72 g            & 6.15 g \\
 ~          & ~             & \cite{Danevich:2005b} & \cite{Belli:2007c}& \cite{Casali:2014a} \\
\hline
 ~          & $^{40}$K      &                       & $<1500$           &  \\
\hline
 ~          & $^{60}$Co     &                       & $<26$             &  \\
\hline
 ~          & $^{137}$Cs    &                       & $<81$             &  \\
\hline
 ~          & $^{147}$Sm    &                       & ~                 & 454  \\
\hline
 ~          & $^{207}$Bi    &                       & $<9$              &  \\
\hline
 ~          & $^{152}$Eu    &                       & $949(48)$         &  \\
\hline
 ~          & $^{154}$Eu    &                       & $212(35)$         &  \\
\hline
 $^{232}$Th & $^{232}$Th    &                       &                   & 3.5  \\
\hline
 ~          & $^{228}$Ac    &                       & $<200$            &  \\
\hline
 ~          & $^{212}$Pb    &                       & $<250$            &  \\
\hline
 ~          & $^{208}$Tl    &                       & $<130$            &  \\
\hline
 $^{238}$U  & $^{226}$Ra    &                       &                   & 2.9 \\
 \hline
 ~          & $^{214}$Po    &                       & $<170$            & \\
 \hline
 ~          & $^{214}$Bi    &                       & $<70$             & \\
 \hline
 ~          & $^{210}$Po    &                       &                   & 6.2 \\
\hline
 \multicolumn{2}{|c|} {Total $\alpha$ activity} & $<10$ &               &   \\
 \hline
 \end{tabular}
 \end{center}
 \label{tab5}
 \end{table}

\noindent
\begin{table}[ht]
\caption{Radioactive contamination of ZnSe and Zn$^{82}$Se crystal
scintillators.}
\begin{center}
\begin{tabular}{|l|l|l|l|l|}
 \hline
 Chain      & Source        & \multicolumn{3}{|c|} {Activity (mBq/kg) in sample, Reference}   \\
 \cline{3-5}
 ~          & ~             &  ZnSe                     & Zn$^{82}$Se                 & Zn$^{82}$Se  \\
 ~          & ~             &  431 g                    & $\approx440$ g, 3 samples   & 174 -- 481 g, 26 samples   \\
 ~          & ~             & \cite{Beeman:2013a}       & \cite{Artusa:2016a}         & \cite{Azzolini:2018}  \\
  \hline
 $^{232}$Th & $^{232}$Th    & 0.0172(46)                & $<0.005-0.013(4)$           & $<0.0005-0.009(1)$ \\
\hline
 ~          & $^{228}$Th    & 0.0111(37)                & $0.022(4)-0.032(7)$         & $0.0023(8)-0.027(2)$ \\
\hline
 $^{238}$U  & $^{238}$U     & 0.0246(55)                & $<0.01-0.020(5)$            & $<0.001-0.013(2)$ \\
\hline
 ~          & $^{234}$U     & 0.0178(33)                & ~                           & $0.001(2)-0.015(4)$ \\
 \hline
 ~          & $^{234}$U$+^{226}$Ra  & 0.0178(33)        & $0.023(5)-0.042(7)$         &  \\
\hline
 ~          & $^{230}$Th    & 0.0246(55)                & $0.017(4)-0.019(5)$         & $<0.002-0.016(2)$ \\
\hline
 ~          & $^{226}$Ra    & 0.0178(33)                &  ~                          & $0.004(1)-0.018(2)$ \\
 \hline
 ~          & $^{218}$Po    &                           & $0.020(5)-0.024(5)$         & $0.003(1)-0.020(2)$ \\
\hline
 ~          & $^{210}$Pb    &                           & $0.100(12)-0.250(17)$       & \\
 \hline
 ~          & $^{210}$Po    & $<0.1$                    &  ~                          &  \\
\hline
\end{tabular}
 \end{center}
 \label{tab6}
 \end{table}

\clearpage
 \begin{landscape}
\begin{table*}[htbp]
\caption{Radioactive contamination of NaI(Tl) crystal
scintillators.}
\begin{center}
\resizebox{1.45\textwidth}{!}{
\begin{tabular}{|l|l|l|l|l|l|l|l|l|l|}
 \hline
  Chain      & Source        & \multicolumn{8}{|c|} {Activity (mBq/kg) in sample, Reference}                                                     \\
 \cline{3-10}
 ~          & ~             & 3.5 kg                    & 10.7 kg, \cite{Amare:2006a,Cuesta:2014a}              & 9.7 kg $\times25$     & $\oslash 72\times72$ mm           & $3.4-9.6$ kg                              & 8.47 kg               & 2 kg                & 8 -- 18 kg            \\
 ~          & ~             & \cite{Barton:2000a}       & 12.5 kg, \cite{Amare:2014a,Amare:2015a,Amare:2016a}   & \cite{Bernabei:2008c} &\cite{Fushimi:2014a,Fushimi:2016a} & \cite{Lee:2015a,Kim:2015b,Adhikari:2016a} & \cite{Cherwinka:2014a}&\cite{DAngelo:2016a} &\cite{Adhikari:2018}   \\
    ~       & ~             & UK DM                     & ANAIS                                                 & DAMA/LIBRA            & PICO-LON                          & KIMS-NaI                                  & DM-Ice17              & SABRE               & COSINE-100            \\
 \hline
    ~       & $^3$H         & ~                         & 0.2                                                   & $<0.09$               &                                   &                                           &                       &                     &                       \\
 \hline
  ~         & $^{22}$Na     & ~                         & ~                                                     & $<0.015$              &                                   &                                           &                       &                     &                       \\
 \hline
  ~         & $^{24}$Na     & ~                         & ~                                                     & $<0.00026$            &                                   &                                           &                       &                     &                       \\
\hline
 \end{tabular}
 }
 \end{center}
 \label{tab7}
 \end{table*}
\end{landscape}

\noindent
\begin{table}[ht]
\caption{Radioactive contamination of CsI(Tl) and CsI(Na) crystal
scintillators.}
\begin{center}
\begin{tabular}{|l|l|l|l|l|l|}
\hline
 Chain      & Source        & \multicolumn{4}{|c|} {Activity (mBq/kg) in sample, Reference}                                                     \\
 \cline{3-6}
 ~          & ~             & \multicolumn{3}{|c|} {CsI(Tl)}                            & CsI(Na) \\
 \cline{3-6}
 ~          & ~             & 2 kg $\times 93$  & 6.6 kg            & 4.55 kg, 3 samples & $\approx2$ kg \\
 ~          & ~             & \cite{Zhu:2006a}  &\cite{Lee:2007b}   & \cite{Liu:2015a}  & \cite{Collar:2015a} \\
    ~       & ~             & TEXONO            & KIMS              & China Dark matter Experiment & \\
 \hline
  ~         & $^{40}$K      & ~                 & ~                 & ~ & 17(16) \\
 \hline
  ~         & $^{87}$Rb     & ~                 & 1.2(4)            & ~ & 61 \\
 \hline
  ~         & $^{134}$Cs    &                   & 6.3(7)            &  & 26(2) \\
 \hline
  ~         & $^{137}$Cs    & 61(2)             & 14.1(11)          & $72.9(1)-86.6(1)$ & 28(3) \\
 \hline
 $^{232}$Th & $^{232}$Th    & 0.0091(2)         & 0.0015(3)         & & $<4$ \\
 \hline
 ~          & $^{228}$Th    & ~                 & ~                 & $0.017(2)-0.069(3)$ & \\
  \hline
 $^{235}$U  & $^{235}$U     & $<0.0016$         & ~                 & & \\
 \hline
 ~          & $^{227}$Ac    & ~                 & ~                 & $<(0.2-1.1)\times10^{-3}$ & \\
 \hline
 $^{238}$U  & $^{238}$U     & 0.0101(2)         & 0.009(3)          & & $<12$ \\
 \hline
 ~          & $^{226}$Ra    & ~                 & ~                 & $0.022(2)-0.087(3)$ & \\
\hline
 \end{tabular}
 \end{center}
 \label{tab8}
 \end{table}

\noindent
\begin{table}[ht]
\caption{Radioactive contamination of LiI(Eu), SrI$_2$(Eu),
LaCl$_3$(Ce), CeBr$_3$, Cs$_2$HfCl$_6$ and YVO$_4$ crystal
scintillators.}
\begin{center}
\begin{tabular}{|l|l|l|l|l|l|l|l|}
  \hline
 Chain      & Source        & \multicolumn{6}{|c|} {Activity (mBq/kg) in sample, Reference}   \\
 \cline{3-8}
 ~          & ~             & LiI(Eu)           & SrI$_2$(Eu)               &   LaCl$_3$(Ce)        & CeBr$_3$              & Cs$_2$HfCl$_6$         & YVO$_4$\\
 \cline{3-8}
 ~          & ~             & 26 g              & $\oslash 13\times11$ mm   &  49.7 g               & 222 g                 & 12 g                   & 22 g \\
 ~          & ~             & 2 samples         & 6.6 g                     & \cite{Bernabei:2005a} & \cite{Lutter:2013a}   & \cite{Cardenas:2017a}  & \cite{Pattavina:2018} \\
 ~          & ~             & \cite{Belli:2013a}& \cite{Belli:2012b}        &                       &                       &                        & \\
 \hline
  ~         & $^{40}$K      & $<177$            & $<200$                    &                       & $<1.9$                & $<0.18$                & $<220$ \\
 \hline
  ~         & $^{60}$Co     & ~                 & ~                         &                       & 1.4(4)                & $<19$                  & $<2.2$ \\
  \hline
  ~         & $^{82}$Br     & ~                 & ~                         &                       & 18(4)                 & ~                      & \\
\hline
 \end{tabular}
 \end{center}
 \label{tab9}
 \end{table}

\clearpage
\begin{table}[ht]
\caption{Radioactive contamination of Gd$_2$SiO$_5$ (GSO) and
Bi$_4$Ge$_3$O$_{12}$ (BGO) crystal scintillators.}
\begin{center}
\begin{tabular}{|l|l|l|l|l|l|l|}
 \hline
 Chain      & Source        & \multicolumn{5}{|c|} {Activity (mBq/kg) in sample, Reference}   \\
 \cline{3-7}
 ~          & ~             & \multicolumn{2}{|c|} {GSO}                                    & \multicolumn{3}{|c|} {BGO} \\
 \cline{3-7}
 ~          & ~             & 635 g                         & 1.744 kg                      & $0.28-0.49$ kg                        & 46 g                          & Table 1  \\
 ~          & ~             & \cite{Danevich:2001a}         & \cite{Wang:2002a}             & \cite{Balysh:1993a}                   & \cite{Marcillac:2003a}        & in \cite{Coron:2008a} \\
 \hline
  ~         & $^{40}$K      & $<14$                         &                               &                                       &                               & $<7$ \\
\hline
  ~         & $^{138}$La    & $<55$                         &                               &                                       &                               &  \\
\hline
  ~         & $^{147}$Sm    &                               & $\approx100$                  &                                       &                               &  \\
\hline
  ~         & $^{207}$Bi    &                               &                               & $520-930$                             & 3000                          & $7(2)-4400$ \\
\hline
 \end{tabular}
 \end{center}
 \label{tab10}
 \end{table}

\clearpage
 \begin{landscape}
\begin{table*}[htbp]
\caption{Radioactive contamination of molybdate crystal
scintillators.}
\begin{center}
\resizebox{1.5\textwidth}{!}{
\begin{tabular}{|l|l|l|l|l|l|l|l|l|}
 \hline
 Chain      & Source        & \multicolumn{7}{|c|} {Activity (mBq/kg) in sample, Reference}   \\
 \cline{3-9}
 ~          & ~             & Li$_2$MoO$_4$                         & Li$_2$$^{100}$MoO$_4$                 & Li$_2$Mg$_2$(MoO$_4$)$_3$         & ZnMoO$_4$                                 & Zn$^{100}$MoO$_4$                         & CaMoO$_4$                         & $^{48depl}$Ca$^{100}$MoO$_4$ \\
 \cline{3-9}
 ~          & ~             & $151-242$ g, \cite{Armengaud:2017a}   & $186-204$ g, \cite{Armengaud:2017a}   & 10.2 g, \cite{Danevich:2018a}     & $\approx 340$ g, \cite{Armengaud:2017a}   & $\approx 380$ g, \cite{Armengaud:2017a}   & $82-100$ g, \cite{Annenkov:2008a} & $\approx310$ g, \cite{Luqman:2017a} \\
 \hline
  ~         & $^{40}$K      & $<3.2-62(2)$                          & $<3.5$                                &                                   &                                           &                                           & $<1.1$                            &   \\
 \hline
 ~          & $^{90}$Sr     &                                       &                                       &                                   &                                           &                                           & $<23$                             &   \\
\hline
  ~         & $^{190}$Pt    & $<0.018$                              & $<0.003$                              &                                   & 0.0026(13)                                &                                           &                                   &    \\
 \hline
 $^{232}$Th & $^{232}$Th    & $<0.018$                              & $<0.003$                              & $<0.95$                           & $<0.0014$                                 & $<0.008$                                  & $<0.7$                            & $<0.05$   \\
\hline
 \end{tabular}
 }
 \end{center}
 \label{tab11}
 \end{table*}
\end{landscape}

\clearpage
 \begin{landscape}
\begin{table*}[htbp]
\caption{Radioactive contamination of cadmium tungstate (CdWO$_4$)
crystal scintillators.}
\begin{center}
\resizebox{1.2\textwidth}{!}{
\begin{tabular}{|l|l|l|l|l|l|l|l|l|}
\hline
 Chain      & Source        & \multicolumn{7}{|c|} {Activity (mBq/kg) in sample, Reference}   \\
 \cline{3-9}
 ~          & ~             & \multicolumn{4}{|c|} {CdWO$_4$}                                                                               & $^{106}$CdWO$_4$              & \multicolumn{2}{|c|} {$^{116}$CdWO$_4$} \\
 \cline{3-9}
 ~          & ~             & 72 g                  & 1046 g                                & 434 g             & 496 g                     & 215 g                         & 330 g                                 & 580 g, 582 g  \\
 ~          & ~             &\cite{Georgadze:1996a} & \cite{Danevich:1996b,Burachas:1996a}  &\cite{Belli:2007a} & \cite{Arnaboldi:2010a}    & \cite{Belli:2010a,Belli:2012a} & \cite{Danevich:2003a,Danevich:2003b}  & \cite{Poda:2013a,Barabash:2016a,Polischuk:2017a} \\
 \hline
  ~         & $^{40}$K      & 3.6(20)               & $<1.7$                                & $<5$              &                           & $<1.4$                        & 0.3(1)                                & $<0.2$  \\
 \hline
  ~         & $^{90}$Sr     & $<1.4$                & $<3$                                  &                   &                           &                               & $<0.2$                                &    \\
\hline
  ~         & $^{90}$Sr$-^{90}$Y &                  &                                       & $<1$              &                           & $<0.3$                        &                                       &    \\
\hline
  ~         & $^{110m}$Ag   &                       &                                       &                   &                           &                               &                                       & 0.12(4)   \\
\end{tabular}
 }
 \end{center}
 \label{tab12}
 \end{table*}
\end{landscape}

\clearpage
 \begin{landscape}
\begin{table*}[htbp]
\caption{Radioactive contamination of magnesium (MgWO$_4$),
calcium (CaWO$_4$), zinc (ZnWO$_4$) and lead (PbWO$_4$) tungstate
crystal scintillators.}
\begin{center}
\resizebox{1.5\textwidth}{!}{
\begin{tabular}{|l|l|l|l|l|l|l|l|l|l|l|}
 \hline
 Chain      & Source        & \multicolumn{9}{|c|} {Activity (mBq/kg) in sample, Reference}   \\
 \cline{3-11}
 ~          & ~             & MgWO$_4$                  & \multicolumn{4}{|c|} {CaWO$_4$}                                                                           & \multicolumn{2}{|c|} {ZnWO$_4$}               & \multicolumn{2}{|c|} {PbWO$_4$} \\
 \cline{3-11}
 ~          & ~             & 0.9 g                     & 54 g                  & 189 g                 & $328-740$ g           &$155-310$ g                        & 163 g                 & $117-699$ g           & 182 g                 & 454 g$^b$ \\
 ~          & ~             & \cite{Danevich:2009a}     & \cite{Cebrian:2004a}  & \cite{Zdesenko:2005a} & \cite{Danevich:2011a} & \cite{Munster:2014a,Strauss:2015a}& \cite{Bavykina:2009a} & \cite{Belli:2011c}    & \cite{Danevich:2006a} & \cite{Beeman:2013b} \\
 \hline
  ~         & $^{40}$K      & $<1.6\times10^3$          & $<100$                & $<12$                 & $<21$                 & $<20$                             & $<321$                & $<0.02$               &                       & \\
 \hline
  ~         & $^{65}$Zn     & ~                         & ~                     &                       &                       &                                   &                       & $0.5(1)-0.8(2)$       &                       & \\
 \hline
 ~          & $^{90}$Sr$-^{90}$Y & $<1.5\times10^3$     & ~                     & $<70$                 & $<26$                 &                                   &                       & $<0.1 $               &                       & \\
 \hline
  ~         & $^{137}$Cs    & ~                         & $<10$                 & $<0.8$                &                       & $<1.2$                            &                       & $<0.05$               &                       & \\
\end{tabular}
 }
 \end{center}
 \label{tab13}
 \end{table*}
\end{landscape}

Radioactive contamination of scintillators is summarized in Table
\ref{tab14} where the data for germanium crystals used in
ultra-low background HP Ge detector
\cite{Klapdor:2002a,Dorr:2003a} are given for comparison. As one
can see in Tables \ref{tab3}-\ref{tab13}, authors present their
data for different members of the $^{232}$Th and $^{238}$U
families. In most cases the measured nuclides are short living
daughters of $^{226}$Ra and $^{228}$Th. Therefore, in Table
\ref{tab10} we present data for $^{228}$Th and $^{226}$Ra.
Activity of $^{210}$Po is typically measured in scintillators,
while its origin is mostly contamination by lead ($^{210}$Pb). For
this reason data on $^{210}$Pb activity is presented in Table
\ref{tab14}. However, one should keep in mind that equilibrium
between $^{210}$Pb and its daughter $^{210}$Po is broken as usual,
as it was discussed above.

\clearpage
 \begin{landscape}
\begin{table*}[htbp]
\caption{Radioactive contamination of scintillators (mBq/kg).
 Data for germanium crystals of HP Ge detectors \cite{Klapdor:2002a,Dorr:2003a} are given for comparison.}
\begin{center}
\resizebox{1.15\textwidth}{!}{
\begin{tabular}{|l|l|l|l|l|l|l|}
 \hline
 Scintillator           & $^{40}$K          & $^{228}$Th                & $^{226}$Ra            & $^{210}$Pb            &  Total $\alpha$           & Particular radioactivity                      \\ % & References \\
 ~                      &                   & ~                         & ~                     & ~                     & (U + Th)                  & ~                                             \\ % & ~ \\
 \hline
 Plastic scintillator   & 2                 & $<0.0001-9$               & $0.09-1$              &                       &                           & $^{14}$C                                      \\ % & \cite{Argyriades:2010b,Heusser:2015a} \\
\hline
 Liquid scintillator    & $<1\times10^{-7}-3\times10^{-5}$  & $(<4\times10^{-9}-3\times10^{-7})$$^a$  & $(<10^{-9}-6\times10^{-8})$$^b$  & $2\times10^{-6}-0.06$ & ~  & $^{14}$C                    \\ % & \cite{Zuzel:2015a,Keefer:2011a} \\
 \hline
 LiF(W)                 & $<5$              & $<0.6$                    & $3$                   &                       &                           & $^{180}$W                                     \\ % & \cite{Belli:2012c} \\
\hline
 CaF$_2$(Eu)            & $<7$              & $0.13-41$                 & $0.05-75$             & 0.9                   & 8                         & $^{48}$Ca, $^{151}$Eu                         \\ % & \cite{Belli:2007b,Ogawa:2003a} \\
\hline
 SrF$_2$                &                   & 21                        & 496                   & 72$^c$                &                           &                                               \\ % & \cite{Diaz:2013a} \\
\hline
 CeF$_3$                & $<330$            & 1010                      & $<60$                 & $<280$$^c$            &                           &                                               \\ % & \cite{Belli:2003a} \\
\hline
 BaF$_2$                & ~                 & $1.4\times10^3$           & $(1.4-7.8)\times10^3$ & 990                   &                           &                                               \\ % & \cite{Belli:2014a} \\
 \hline
 YAG(Nd)                &                   &                           &                       &                       & $<20$                     & $^{144}$Nd, $^{150}$Nd                        \\ % & \cite{Danevich:2005b} \\
\hline
 Li$_6$Eu(BO$_3)_3$     & $<1500$           & 3.5$^a$                   & 2.9                   & 6.2                   &                           & $^{151}$Eu, $^{152}$Eu, $^{154}$Eu            \\ % & \cite{Belli:2007c,Casali:2014a} \\
\hline
 ZnSe                   & ~                 &$<0.0004-0.02$$^a$         & $<0.0004-0.03$$^b$    &  $<0.1$$^c$           &                           & $^{65}$Zn, $^{75}$Se, $^{82}$Se               \\ % & \cite{Beeman:2013c,Arnaboldi:2011a} \\
\hline
 Zn$^{82}$Se            & ~                 & $0.02-0.03$               &  $0.02-0.04$$^d$      & $0.1-0.3$             &                           & $^{82}$Se                                     \\ % & \cite{Artusa:2016a} \\
\hline
 LaCl$_3$(Ce)           & ~                 & $<0.4$                    & $<40$                 &                       &                           & $^{138}$La, $^{235}$U                         \\ % & \cite{Bernabei:2005a} \\
\hline
 Cs$_2$HfCl$_6$         & $<0.18$           & $<6$                      &                       &                       &                           & $^{174}$Hf                                    \\ % & \cite{Cardenas:2017a} \\
\hline
 LiI(Eu)                & $<180$            & $<0.4$                    & $<1.1$                & $<2$$^c$              &                           & $^{151}$Eu                                    \\ % & \cite{Belli:2013a} \\
\hline
 NaI(Tl)                & $0.3-81$          & $0.0008-0.18$             & $<0.0002-1$           & $0.02-10$             & $0.08-3$                  & $^3$H, $^{129}$I                              \\ % & \cite{Barton:2000a,Amare:2006a,Cuesta:2014a,Amare:2014a,Amare:2015a,Amare:2016a,Bernabei:2008c,Fushimi:2014a,Fushimi:2016a,Lee:2015a,Kim:2015b,Adhikari:2016a,Cherwinka:2014a,DAngelo:2016a} \\
\hline
 SrI$_2$(Eu)            & $<200$            & 6                         & 100                   & $<180$                &  ~                        & $^{151}$Eu                                    \\ % & \cite{Belli:2012b} \\
\hline
 CsI(Na)                & 17                & $<0.4$$^a$                & $<12$$^b$             & ~                     &  ~                        & $^{134}$Cs, $^{137}$Cs                        \\ % & \cite{Collar:2015a} \\
\hline
 CsI(Tl)                & ~                 & $(0.009^a-0.07)$          & $(0.009^b-0.09)$      &                       &                           & $^{134}$Cs, $^{137}$Cs                        \\ % & \cite{Zhu:2006a,Lee:2007b,Liu:2015a} \\
\hline
 CeBr$_3$               & $<1.9$            & $<2$                      & $<0.5$                & $<600$                &                           & $^{82}$Br, $^{139}$Ce                         \\ % & \cite{Lutter:2013a} \\
\hline
 GSO(Ce)                & $<14$             & $2.3-107$                 & $0.27$                & 200                   & $40-220$                  & $^{152}$Gd                                    \\ % & \cite{Danevich:2001a,Wang:2002a} \\
\hline
 BGO                    & $<7$              & 6                         &                       &                       &                           & $^{209}$Bi, $^{207}$Bi                        \\ % & \cite{Balysh:1993a,Marcillac:2003a,Coron:2008a} \\
 \hline
 Li$_2$MoO$_4$          & $<3-60$           & $<0.018$                  & $<0.04-0.13$          & $(0.08-0.2)$$^c$      & ~                         & $^{100}$Mo                                    \\ % & \cite{Armengaud:2017a} \\
 \hline
 Li$_2$$^{100}$MoO$_4$  & $<3$              & $<0.006$                  & $<0.007$              & $(0.06-0.23)$$^c$     & ~                         & $^{100}$Mo                                    \\ % & \cite{Armengaud:2017a} \\
\hline
 Li$_2$Mg$_2$(MoO$_4)_3$&~                  & $<1.1$                    & $<1.7$                & 6$^c$                 & ~                         & $^{100}$Mo                                    \\ % & \cite{Danevich:2018a} \\
\hline
 ZnMoO$_4$              & ~                 & $<0.005$                  & $<0.006$              & 0.6-1.3$^c$           & ~                         & $^{100}$Mo                                    \\ % & \cite{Armengaud:2017a} \\
\hline
 Zn$^{100}$MoO$_4$      & ~                 & $<0.008$                  & $0.014-0.023$         & $0.8-2.4$$^c$         &                           & $^{100}$Mo                                    \\ % & \cite{Armengaud:2017a} \\
\hline
 CaMoO$_4$              & $<1.1$            & $0.04-0.4$                & $0.13-2.5$            & $<8-550$$^c$          &                           & $^{48}$Ca, $^{100}$Mo                         \\ % & \cite{Annenkov:2008a} \\
\hline
 $^{48depl}$Ca$^{100}$MoO$_4$& ~              & $<0.05^a$                 & $0.07$                & 7                     & ~                         & $^{48}$Ca, $^{100}$Mo                         \\ % & \cite{Luqman:2017a} \\
%\hline
% PbMoO$_4$             &                   &                           &                       & $(67-192)\times10^3$  &                           & $^{210}$Pb, $^{100}$Mo                        \\ % & \cite{Zdesenko:1996a} \\
\hline
 MgWO$_4$               & $<1.6\times10^3$  & $<50$                     & $<50$                 & $5.7\times10^3$$^c$   & $5.7\times10^3$           & $^{180}$W                                     \\ % & \cite{Danevich:2009a} \\
 \hline
 CaWO$_4$               & $<12$             & $0.6$                     & $0.04-107$            & $<190-4800$           & $1-1400$                  & $^{48}$Ca, $^{180}$W                          \\ % & \cite{Cebrian:2004a,Zdesenko:2005a,Danevich:2011a} \\
 \hline
 ZnWO$_4$               & $<0.02$           & $0.002-0.018$             & $0.002-0.025$         & $<0.01^c$             & $0.18-2.3$                & $^{65}$Zn, $^{180}$W                          \\ % & \cite{Danevich:2005a,Bavykina:2009a,Belli:2011c} \\
 \hline
 CdWO$_4$               & $<1.7-4$          & $<0.003-0.015$            & $<0.007$              & $<0.06^c$             & 0.26                      & $^{113}$Cd, $^{116}$Cd, $^{180}$W             \\ % & \cite{Georgadze:1996a,Danevich:1996b,Belli:2007a} \\
 \hline
 $^{106}$CdWO$_4$       & $<1.4$            & 0.042                     & 0.012                 & $<0.2^c$              & 2.1                       & $^{113m}$Cd, $^{113}$Cd, $^{116}$Cd, $^{180}$W \\ % & \cite{Belli:2012a} \\
 \hline
 $^{116}$CdWO$_4$       & $<0.2-0.3$        & $0.02-0.07$               & $<0.004$              & 0.23$^c$              & $1.4-2.9$                 & $^{113m}$Cd, $^{113}$Cd, $^{116}$Cd, $^{180}$W \\ % & \cite{Danevich:2003a,Poda:2013a,Barabash:2016a,Polischuk:2017a} \\
 \hline
 PbWO$_4$               & ~                 & $<13$                     & $<10$                 & $(5-8)\times10^4$$^c$ & ~                         & $^{210}$Pb, $^{180}$W                         \\ % & \cite{Danevich:2006a} \\
 \hline
 PbWO$_4^e$             & ~                 & $0.05^a$                  & 1.4                   & 190$^c$               & ~                         & $^{210}$Pb, $^{180}$W                         \\ % & \cite{Beeman:2013b} \\
\hline
 HP Ge                  & ~                 & $<2\times10^{-5}$$^a$     & $<2\times10^{-5}$$^b$ &                       &                           & $^{76}$Ge                                     \\ % & \cite{Klapdor:2002a,Dorr:2003a} \\
\hline
 \multicolumn{7}{l}{$^{a}$~Activity of $^{232}$Th.}\\
 \multicolumn{7}{l}{$^{b}$~Activity of $^{238}$U.}\\
 \multicolumn{7}{l}{$^{c}$~Activity of $^{210}$Po.}\\
 \multicolumn{7}{l}{$^{d}$~Activity of $^{238}$U$+^{226}$Ra.}\\
 \multicolumn{7}{l}{$^{e}$~Produced from ancient lead.}\\
 \end{tabular}
 }
 \end{center}
 \label{tab14}
 \end{table*}
\end{landscape}

Liquid scintillators are the most radiopure scintillation
materials with a nBq/kg~--~pBq/kg radiopurity level of $^{40}$K, U
and Th \cite{Zuzel:2015a,Keefer:2011a}. However, cosmogenic
$^{14}$C remains the main source of the background counting rate
($0.4-0.7$ mBq/kg, at energies below 0.25 MeV) of large
low-background liquid scintillation detectors, despite the
significant efforts to reduce its concentration. Radioactive
contamination of plastic scintillators is significantly higher. It
can be explained by polymerization process and mechanical
treatment of the material, as well as by absence of a strong
motivation to obtain highly radiopure scintillation material
similar to the Borexino and KamLAND experiments. Crystal
scintillators like ZnWO$_4$, CdWO$_4$ (including produced from
enriched $^{106}$Cd and $^{116}$Cd), Li$_2$MoO$_4$, ZnMoO$_4$
(including enriched in $^{100}$Mo), ZnSe (including enriched in
$^{82}$Se), specially developed for low background experiments
CaF$_2$(Eu), NaI(Tl), CsI(Tl) and CaMoO$_4$ have rather low
contamination by $^{226}$Ra and $^{228}$Th on the level of $\sim
0.001-0.1$ mBq/kg.

\begin{figure}
\begin{center}
\resizebox{0.65\textwidth}{!}{\includegraphics{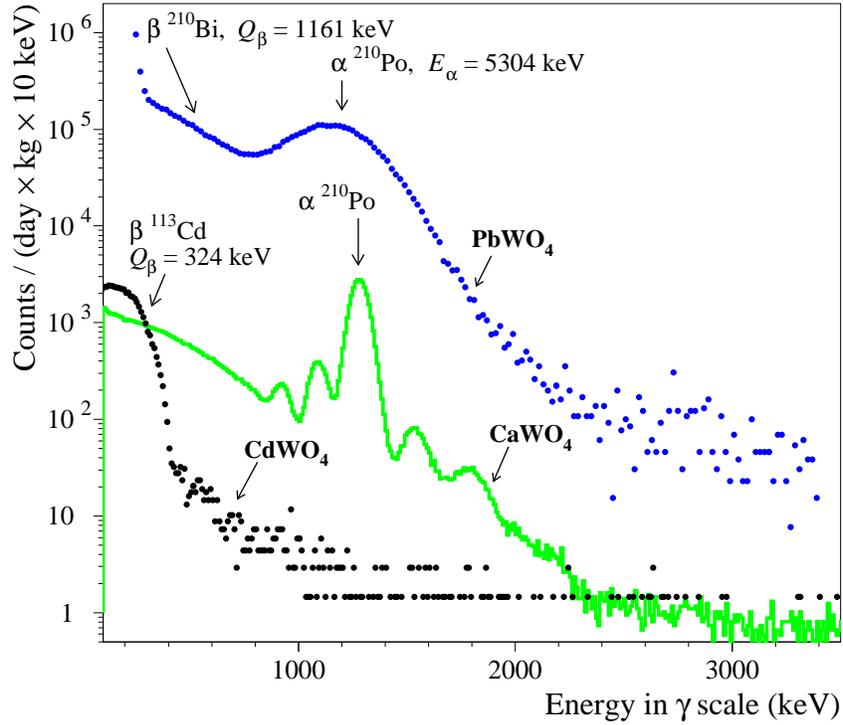}}
\caption{Energy spectra of CaWO$_4$ (189 g, 1734 h), CdWO$_4$ (448
g, 37 h), and PbWO$_4$ (185 g, 2.15 h) scintillation crystals
measured in the low background set-up in the Solotvina Underground
Laboratory (the PbWO$_4$ crystal was measured without shield). The
CaWO$_4$ crystal is considerably polluted by radionuclides from U
and Th chains \cite{Zdesenko:2005b}. Beta decay of $^{113}$Cd
($T_{1/2}=8.04\times10^{15}$ years) dominates in the low energy
part of the CdWO$_4$ spectrum \cite{Danevich:1996a,Belli:2007a}.
PbWO$_4$ crystal is contaminated by $^{210}$Pb
\cite{Danevich:2006a}.}
 \label{fig:fig12}
\end{center}
\end{figure}

\begin{figure}
\begin{center}
\resizebox{0.65\textwidth}{!}{\includegraphics{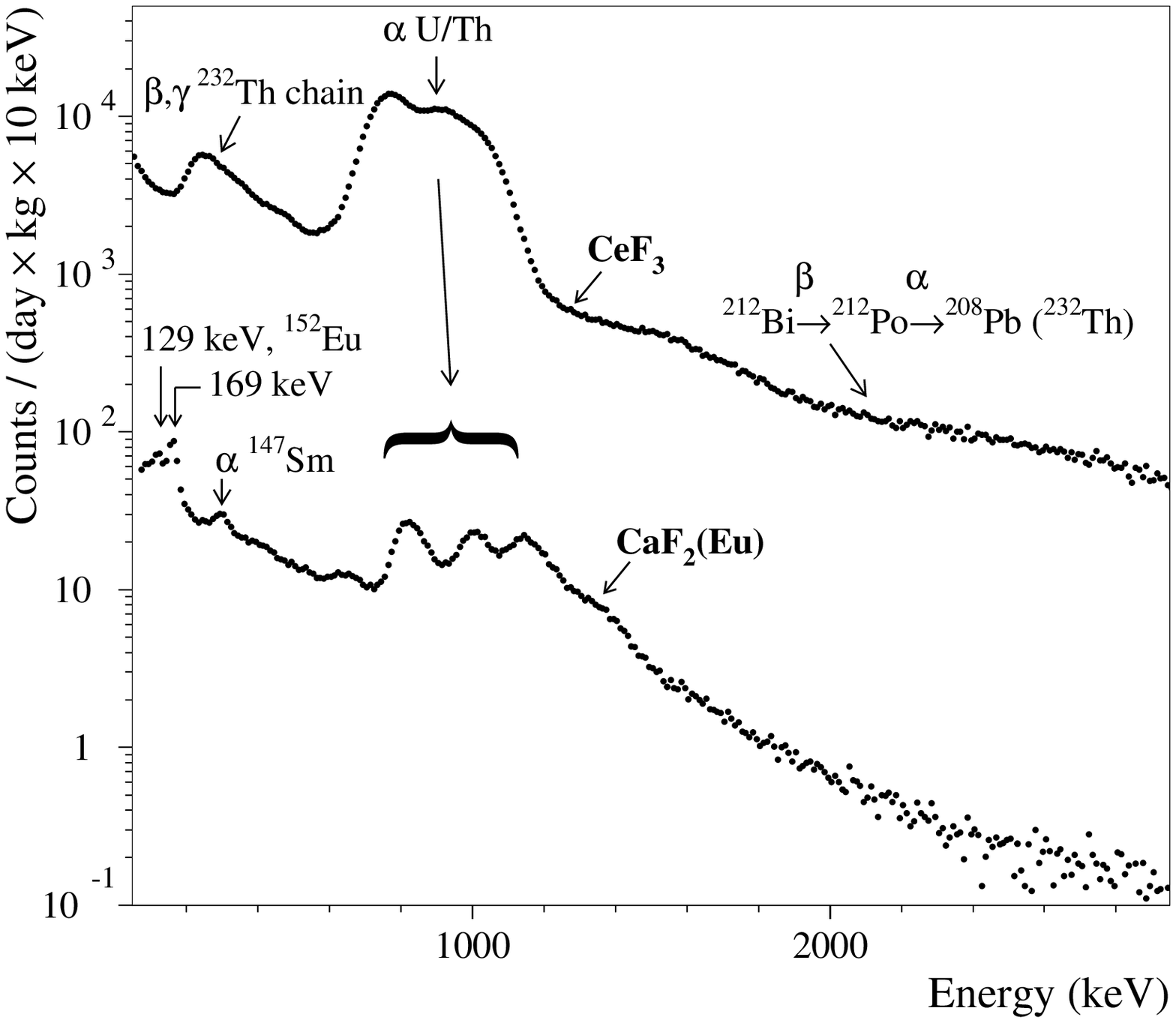}}
\caption{Energy spectra of CeF$_3$ (49 g, 2142 h) and CaF$_2$(Eu)
(370 g, 7426 h) scintillation crystals measured in the DAMA R\&D
low background set-up in the Gran Sasso underground laboratory of
I.N.F.N. (Italy). The CeF$_3$ crystal is considerably polluted by
radionuclides from Th and U chains on the $\sim$ Bq/kg level (see
for details \cite{Belli:2003a}). The background of the CaF$_2$(Eu)
detector is two orders of magnitude lower, and caused by $\alpha$
activity of U/Th daughters on the level of a few mBq/kg
\cite{Belli:2007b}. In the low energy part of the spectrum
accumulated with the CaF$_2$(Eu) crystal one can see peaks from
cosmogenic (or / and neutron induced) $^{152}$Eu ($\approx10$
mBq/kg), as well as $\alpha$ peak of $^{147}$Sm presented in the
crystal with an activity of $\approx0.3$ mBq/kg.}
 \label{fig:fig13}
\end{center}
\end{figure}

A level of crystal scintillators radiopurity is determined first
of all by their chemical composition. For instance, CdWO$_4$ and
ZnWO$_4$ crystals always show low level of internal activity,
while CaWO$_4$ has much higher level of radioactive contamination.
Fig. \ref{fig:fig12} demonstrates the large difference in
radiopurity of CaWO$_4$, CdWO$_4$ and PbWO$_4$. Scintillators
containing rare earth elements (GSO, CeF$_3$) have much higher
level of radioactive trace pollution too. It is due to source of
rare earth mining: they usually are extracted from monazites --
minerals containing a few percents of uranium and thorium. Energy
spectra of CaF$_2$(Eu) and CeF$_3$ crystal scintillators measured
in the same conditions of the DAMA R\&D low background set-up in
the Gran Sasso underground laboratory of I.N.F.N. (Italy) are
presented in Fig. \ref{fig:fig13}. The background counting rate of
the CeF$_3$ detector is two orders of magnitude higher due to the
higher contamination by thorium and uranium. Rather high
radioactive contamination of BaF$_2$ by radium ($^{226}$Ra and
$^{228}$Ra) can be explained by similar chemical properties of
barium and radium. It should be stressed that the chemical and
physical processes involved in scintillators production are
sensitive to the chemical properties of materials. Therefore, it
might be more accurate to discuss contamination of materials by
chemical elements, not by certain isotopes.

\nopagebreak

\nopagebreak

Presence of radioactive elements (see Table \ref{tab2}) obviously
determines radioactivity of scintillators like $\beta$ active
$^{113}$Cd in CdWO$_4$, $\alpha$ active $^{152}$Gd in GSO,
$^{138}$La in LaCl$_3$ and LaBr$_3$, $^{176}$Lu in Lu$_2$SiO$_5$
and LuI$_3$. Beta active $^{210}$Pb is usually present in PbWO$_4$
and PbMoO$_4$ (however, this problem can be overcome by producing
lead containing scintillators from archaeological lead
\cite{Alessandrello:1998a,Danevich:2009a,Beeman:2013b}).

Two neutrino $2\beta$ decay, despite it is the rarest decay ever
observed, becomes one of the most significant background sources in
the experiments aiming to search for $0\nu2\beta$ decay with
smaller $Q_{2\beta}$. For this reason the AMoRE collaboration
developed CaMoO$_4$ crystal scintillators enriched in the isotope
$^{100}$Mo and depleted in $^{48}$Ca
\cite{Bhang:2012a,AMoRE}. Another example is CdMoO$_4$ crystal
scintillators proposed to search for $0\nu2\beta$ decay of
$^{116}$Cd and $^{100}$Mo \cite{Xue:2017}: in this case the
$2\nu2\beta^-$ decays of $^{100}$Mo will generate background in
the region of interest of $^{116}$Cd.

\section{Development of radiopure scintillation materials}
\label{sec:RD}

A strong R\&D is required to obtain a radiopure scintillation
material. There were several systematic programmes to elaborate
radiopure scintillators. Extremely radiopure liquid scintillators
were developed for the Borexino \cite{Benziger:2008a} and KamLAND
\cite{Suekane:2004a} neutrino experiments. Two factors which
determine success of the projects are as following: 1) organic
materials in principle are much less contaminated in comparison to
inorganic materials; 2) liquids can be effectively purified by
distillation, nitrogen purging, water extraction and/or column
purification \cite{Benziger:2008a,Miramonti:2017a}.

Very low background NaI(Tl) scintillators have been produced by
Saint-Gobain for the DAMA/LIBRA dark matter experiment
\cite{Bernabei:2008c,Bernabei:2016a,Bernabei:2017a}; several R\&D
are in progress to develop radiopure NaI(Tl) scintillators
\cite{Amare:2014a,Fushimi:2016a,Lee:2015a,Cherwinka:2014a,DAngelo:2016a}.
Low background CsI(Tl) were developed in the framework of the KIMS
dark matter experiment \cite{Lee:2007b,Zhu:2006a}. Dependence of
radiopurity of CsI(Tl) on crystal growth conditions was studied in
\cite{Zhu:2006a}. Authors reported a clear increase of $^{226}$Ra
and $^{228}$Th activity along the crystals during the crystal
growth process due to increase of the melt contamination. Clear
increase of radioactive contamination along the crystal boule was
observed in $^{116}$CdWO$_4$ crystal \cite{Barabash:2016b}.

Achievement of a high radiopurity level of scintillators needs a
variety of special measures as it was demonstrated by development
of radiopure crystal scintillators for double beta decay
experiments: $^{106}$CdWO$_4$ \cite{Belli:2010a}, $^{116}$CdWO$_4$
\cite{Barabash:2011a}, $^{48depl}$Ca$^{100}$MoO$_4$
\cite{So:2012a,Lee:2016a}, Zn$^{82}$Se \cite{Dafinei:2017a},
Zn$^{100}$MoO$_4$
\cite{Chernyak:2013a,Berge:2014a,Armengaud:2015a,Chernyak:2015a,Armengaud:2017a}
and Li$_2$$^{100}$MoO$_4$
\cite{Bekker:2016a,Grigorieva:2017a,Armengaud:2017a}.

A programme to develop radiopure scintillation materials for
experiments to search for rare processes could comprise the
following steps \cite{Danevich:2008a}:

1. Careful selection and deep purification of raw materials is
supposed to be the most important issue that needs addressing.
Purification of metals (Zn, Cd, Pb, Bi) by vacuum distillation,
zone melting, and filtering are very promising approaches
\cite{Bernabey:2008a,Kovtun:2011a,Boiko:2011a}, while further
study is necessary for the purification of Li, Na, Ca, Se, Cs, Ba,
Ce, Gd, W, Tl. Molybdenum can be deeply purified by combination of
a double sublimation of molybdenum oxide with subsequent
recrystallization in aqueous solutions \cite{Berge:2014a}.

2. Two step re-crystallization, involving inspection and
assessment of the produced scintillators after each step. A
promising result was obtained with the $^{116}$CdWO$_4$ crystal
scintillator by re-crystallization. The radioactive contamination
of a sample of $^{116}$CdWO$_4$ scintillator by thorium was
reduced by one order of magnitude after the second crystallization
procedure \cite{Barabash:2016b}. A larger number of
recrystallization steps is limited by evaporation of components of
melt (e.g. dominant evaporation of cadmium from CdWO$_4$ melt, or
molybdenum from Li$_2$MoO$_4$, CaMoO$_4$, ZnMoO$_4$), which leads
to violation of stoichiometry of the melt. Besides, one could
expect some influence of growing process to radioactive
contamination of crystal scintillators, which effects were never
systematically studied. One could expect that at a certain stage
of such an R\&D, development of special low background growth
facilities will be necessary. An interesting R\&D was performed in
\cite{Fushimi:2016a} aiming at development of highly radiopure
NaI(Tl) crystal scintillators. Authors observed a significant
improvement of the crystal radiopurity level by utilization of
high purity graphite crucible. Investigations of ceramics used in
the growing set-ups was performed in the framework of radiopure
ZnWO$_4$ crystal scintillators R\&D \cite{Belli:2011c}. We would
like to stress that the low-thermal gradient Czochralski crystal
growth method provides high quality large volume radiopure crystal
scintillators \cite{Shlegel:2017a}. The technique is especially
suitable for enriched crystal scintillators production
\cite{Danevich:2012a}.

3.  Screening at all stages through ultra-low background
$\gamma$-spectrometry is needed in the production of compounds for
crystal growing (choice of raw materials, quality control of
purified elements and compounds).

4. All work should be done using highly pure reagents, lab-ware
and water. Careful protection from radon should be foreseen
(especially in the case of scintillators R\&D for dark matter
experiments). All chemistry should be done in clean room
conditions, and, as far as possible, in radon free atmosphere.

5. Cosmogenic activation is expected to be one of the  most
significant sources of background as the low counting technique
improves. Therefore, underground facilities for crystal
scintillators production could be a natural step to obtain highly
radiopure scintillators.

\section{Conclusions}
\label{sec:con}

Scintillation detectors are widely used to search for rare
processes in nuclear and astroparticle physics. Radioactive
contamination of scintillation materials can be measured by
indirect (inductively coupled plasma mass spectrometry and neutron
activation are the most sensitive approaches) and direct (first of
all, by low-background HP Ge $\gamma$ spectrometry) methods.
However, the most sensitive method to measure internal
contamination of a scintillator are low background measurements
when the scintillator operates as a detector. Time-amplitude
analysis allows to detect at the $\mu$Bq/kg level the fast
sub-chains of the $^{232}$Th, $^{235}$U and $^{238}$U families
which are in equilibrium with $^{228}$Th (from the $^{232}$Th
family), $^{226}$Ra ($^{238}$U) and $^{227}$Ac ($^{235}$U). Alpha
active nuclides can be determined by using pulse-shape
discrimination technique. Estimation of $\beta$ active nuclides
can be done by fit of measured energy spectra using Monte Carlo
simulated models of background.

Liquid scintillators are the most radiopure scintillation
materials with the best up to date achieved residual radioactive
contamination level of nBq/kg~--~pBq/kg. Radioactive contamination
of crystal scintillators varies in a wide range. The most
radiopure crystal scintillators are ZnWO$_4$, CdWO$_4$,
Li$_2$MoO$_4$, ZnMoO$_4$, ZnSe and specially developed for dark
matter experiments NaI(Tl) and CsI(Tl) those radioactive
contamination by $^{228}$Th and $^{226}$Ra does not exceed the
level of $\sim 0.01$ mBq/kg. Main sources of internal
radioactivity of scintillators are daughters of U/Th families,
$^{40}$K, radioactive isotopes of elements which are part of a
scintillator composition. Scintillation materials containing rare
earth elements have comparatively high level of U/Th
contamination. Equilibrium of the $^{232}$Th, $^{235}$U and
$^{238}$U chains is usually broken in scintillators.

Next generation $2\beta$ and dark matter experiments call for
extremely low ($\sim0.1-1~\mu$Bq/kg) level of radioactive
contamination of crystal scintillators, that can be obtained in
the framework of special programmes that should include deep
purification of raw materials, careful screening at all the
stages, application of double crystallization. The programmes
should involve use of radiopure reagents, lab-ware, equipment and
installations, production of raw materials, crystal growing and
their storage in radon free atmosphere. Special efforts are
necessary to prevent cosmogenic and neutron activation of
materials. Development of special underground low background
growth facilities may be necessary to reach a further progress in
production of radiopure crystal scintillators. Tests of crystal
scintillators radioactive contamination on the $0.1-1~\mu$Bq/kg
level can be realized by ultra-low background direct measurements
when a scintillator serves as a detector of its internal
radioactivity. The technique of cryogenic scintillating bolometers
provides the highest sensitivity to $\alpha$ active contaminations
of crystal scintillators.

\section*{Acknowledgments}

F.A.~Danevich gratefully acknowledges support from the Jean
d'Alembert fellowship program (project CYGNUS) of the Paris-Saclay
Excellence Initiative, grant number ANR-10-IDEX-0003-02. Authors
were supported in part by the IDEATE International Associated
Laboratory (LIA), and by the project ``Investigation of neutrino
and weak interaction in double beta decay of $^{100}$Mo'' in the
framework of the Programme ``Dnipro'' based on Ukraine-France
Agreement on Cultural, Scientific and Technological Cooperation.

\end{document}